\documentclass[11pt,a4paper]{article}
\usepackage{graphicx} 
\usepackage{xcolor}
\usepackage{a4wide}
\usepackage{amssymb}
\usepackage{amsmath}
\usepackage{booktabs}
\usepackage{cite}
\usepackage{authblk}
\usepackage[margin=2.5cm,nohead]{geometry}
\usepackage{xspace}
\usepackage{enumitem}
\usepackage{subcaption}

\usepackage[hidelinks, colorlinks=true, allcolors=blue]{hyperref}
\usepackage{cleveref}

\newcommand{\appRef}[1]{app.~\ref{#1}\xspace}
\newcommand{\eqRef}[1]{eq.~\eqref{#1}\xspace}
\newcommand{\eqsRef}[1]{eqs.~\eqref{#1}\xspace}

\newcommand{\figRef}[1]{fig.~\ref{#1}\xspace}
\newcommand{\figsRef}[1]{figs.~\ref{#1}\xspace}
\newcommand{\tabRef}[1]{tab.~\ref{#1}\xspace}

\newcommand{\secRef}[1]{sec.~\ref{#1}\xspace}
\newcommand{\secsRef}[1]{secs.~\ref{#1}\xspace}

\newcommand{\comix}{C\protect\scalebox{0.85}{OMIX}\xspace}

\newcommand{\nnlojet}{NNLO\protect\scalebox{0.85}{JET}\xspace}
\newcommand{\topxx}{Top++\xspace}

\newcommand{\sherpa}{S\protect\scalebox{0.85}{HERPA}\xspace}

\newcommand{\GeV}{\ensuremath{\mathrm{GeV}}\xspace}
\newcommand{\TeV}{\ensuremath{\mathrm{TeV}}\xspace}
\newcommand{\pb}{\ensuremath{\mathrm{pb}}\xspace}

\renewcommand{\t}{\ensuremath{\mathrm{t}}\xspace}

\newcommand{\W}{\ensuremath{\mathrm{W}}\xspace}
\newcommand{\Z}{\ensuremath{\mathrm{Z}}\xspace}

\title{\vspace*{-5mm}Summary of Aspen discussions on perturbative uncertainties beyond 7-point variations}

\author[1]{E.~Canonero}
\affil[1]{\small Physics Department, Royal Holloway, University of London, UK}

\author[1]{G.~Cowan}

\author[2]{S.~Forte}
\affil[2]{\small Dipartimento di Fisica, Universit\`a di Milano \& INFN Milano, Via Celoria 16, I-20133 Milano, Italy}

\author[3]{J.~Huston}
\affil[3]{\small Physics and Astronomy Department, Michigan State University, East Lansing, MI 48824 USA}

\author[4]{S.~Jaskiewicz}
\affil[4]{\small Albert Einstein Center for Fundamental Physics, Institut f\"ur Theoretische Physik,
Universit\"at Bern, Sidlerstrasse 5, CH-3012 Bern, Switzerland}

\author[5]{M.~Kraus}
\affil[5]{\small Departamento de F\'{i}sica Te\'{o}rica, Instituto de F\'{i}sica,
Universidad Nacional Aut\'{o}noma de M\'{e}xico, Cd. de M\'{e}xico C.P. 04510, M\'{e}xico}

\author[6]{C.~T.~Preuss}
\affil[6]{\small Institut f\"ur Theoretische Teilchenphysik und Kosmologie, RWTH Aachen University, 52056 Aachen, Germany}

\author[7,8]{D.~Reichelt}
\affil[7]{\small Institute for Particle Physics Phenomenology, Durham University, Durham DH1 3LE, UK}
\affil[8]{\small CERN, Theoretical Physics Department, CH-1211 Geneva 23, Switzerland}

\author[9]{P.~Skands\footnote{peter.skands@monash.edu  (corresponding author)}}
\affil[9]{\small School of Physics \& Astronomy, Monash University, Melbourne VIC-3800, Australia}

\author[10]{R.~Szafron}
\affil[10]{\small Department of Physics, Brookhaven National Laboratory, Upton, New  York, 11973, USA}

\author[11]{Y.~Ulrich}
\affil[11]{\small  Department of Mathematical Sciences, University of Liverpool,  Liverpool, L69 3BX, UK}

\date{}

\begin{document}
\maketitle

\vspace*{-5mm}\begin{abstract}
This writeup elaborates on constructive discussions which took place at the Aspen Center for Physics, Aug -- Sep 2024, during the  program ``Tightening the gap between scattering amplitudes and events at the LHC at higher orders''. Building on already existing  theoretical understanding from the study of scattering amplitudes and resummations at higher orders, we suggest to complement standard 7-point renormalisation- and factorisation-scale variations by additional estimates of missing higher-order terms which are known to be present in the perturbative series. We also summarise a recent proposal for at least one way that non-Gaussian theory uncertainties might be treated within a frequentist framework. This may help to provide more robust and comprehensive uncertainty estimates for experimental and phenomenological applications of perturbative  calculations. 
\end{abstract}
\vfill

\setcounter{tocdepth}{2}
\tableofcontents

\vfill
\clearpage
\section{Introduction}
The credibility of scientific results rests on the reliability of their uncertainty estimates.
In collider physics, uncertainty estimates at the lowest perturbative orders frequently fail to span actual differences between theory and data. This is a problem not only for practical applications (e.g., in the context of experimental studies) but also because it undermines our confidence in one of the main tools of collider physics; we tend not to trust the uncertainties at a given order fully, until  the calculation at the next order has been done.

That comes with a high cost in both human and computing resources. And then often reveals that, indeed, the standard approach to estimating the uncertainty at the preceding order did fall short (see, e.g.,~\cite{Bagnaschi:2014wea} for a survey). In hindsight, we can often claim to understand \emph{why} the approach failed. The question we wish to examine here is whether (some of) that understanding can be quantified, and used to provide more comprehensive and robust perturbative uncertainty estimates. 

The problem is especially acute for applications that, for one reason or another, are forced to operate at low perturbative orders. This includes the majority of present-day Monte-Carlo event generators~(see \cite{Buckley:2011ms,Campbell:2022qmc,vanBeekveld:2026uxl} for reviews), which typically operate at LO or NLO fixed-order (FO) accuracy for the hard process, supplemented by parton showers (PS) that achieve LL or NLL accuracy for soft/collinear all-orders corrections. 

\begin{figure}[t]
\centering
\includegraphics*[width=0.42\textwidth]{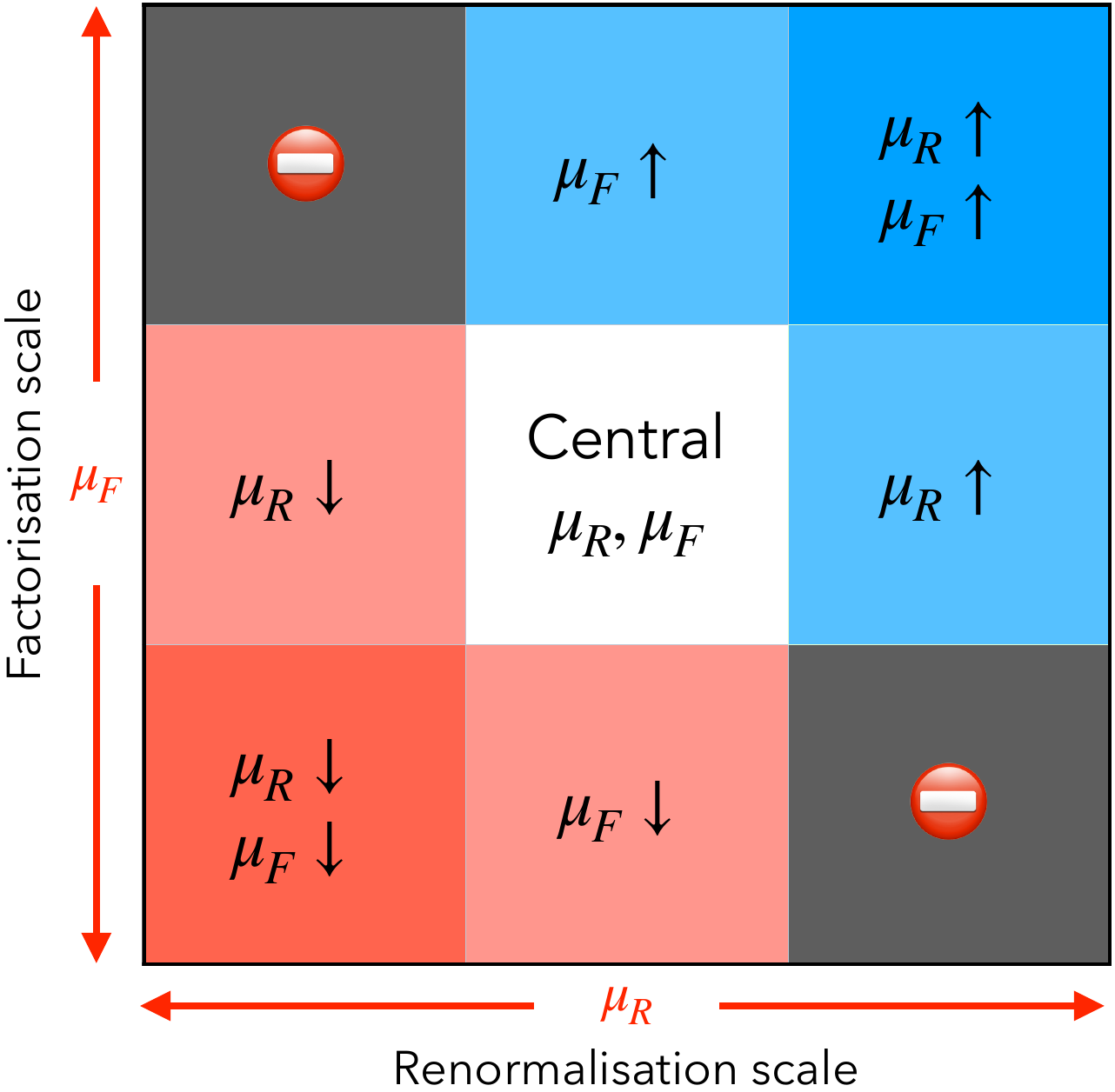}
\caption{Illustration of standard 7-point scale variations, increasing and decreasing the renormalisation and factorisation scales ($\mu_R$ and $\mu_F$ respectively) independently around a single central choice, but excluding the two variations that go in opposite directions. \label{fig:7point}}
\end{figure}

For some (mostly $2\to 1$ and $2\to 2$) processes, NNLO+PS matching has also started to become available~\cite{Hamilton:2012rf,Alioli:2015toa,Mazzitelli:2021mmm,El-Menoufi:2024sys}, and efforts are also underway to develop parton showers that can achieve NNLL accuracy~\cite{vanBeekveld:2024wws}. Nevertheless, at least for the near term (e.g., in the context of current LHC Run-3 data analyses), it remains likely that many studies will still be carried out effectively at LO/NLO + LL/NLL accuracy. For these applications, perturbative uncertainties that are more robust than the conventional so-called 7-point renormalisation- and factorisation-scale variations (illustrated in \figRef{fig:7point}) would be welcome. 


Several proposals have been made for how to go beyond conventional scale variations~\cite{Cacciari:2011ze,David:2013gaa,Bagnaschi:2014wea,NNPDF:2019ubu,Bonvini:2020xeo,Duhr:2021mfd,Ghosh:2022lrf,Tackmann:2024kci,Lim:2024nsk,NNPDF:2024dpb,Gambhir:2025lka,Bonvini:2026feq}, in most cases also including dedicated validation checks against known processes. A common focus of several of these~\cite{Cacciari:2011ze,Bonvini:2020xeo,Duhr:2021mfd,Ghosh:2022lrf,Lim:2024nsk} has been on exploiting information obtained at previous orders and/or in related processes, to make better-informed guesses about the next uncontrolled order for the process at hand, in some cases adopting a Bayesian framework to help quantify what constitutes a ``better informed guess''~\cite{Cacciari:2011ze,Bagnaschi:2014wea,Duhr:2021mfd,Bonvini:2026feq}. 

There is also a growing consensus on the importance of accounting for correlations among missing-higher-order uncertainties, e.g., using covariance matrices~\cite{NNPDF:2019ubu,NNPDF:2024dpb}, explicit theory nuisance parameters~\cite{Tackmann:2024kci,Lim:2024nsk}, and/or well-motivated all-orders completions~\cite{Gambhir:2025lka}.

Here, we take an approach we believe is in line with that of \cite{Tackmann:2024kci}, examining a few well-known examples of higher-order terms that are not accessed by conventional renormalisation- and factorisation-scale variations --- but which are each theoretically well understood and sufficiently universal that it should be possible to formulate a generic framework aimed at estimating their possible sizes, at least within the limited context of making uncertainty estimates. Moreover, since these terms are linearly independent of (and often functionally quite different from) those generated by scale variations, we argue that such variations could be considered on a similar footing as scale variations in the context of comprehensive uncertainty estimates. We shall seek to do so while remaining within a frequentist statistical approach, not because we have any objections to Bayesian methodologies, but because we believe the underlying issues to be sufficiently general that it should be possible to formulate improved approaches for any choice of statistical framework.

In \secRef{sec:beyond}, we summarise some known sources of generic higher-order terms --- some of which can be large --- which standard scale variations are either blind to or cannot be relied upon to provide robust explorations of. In the context of the Aspen workshop, the following three were studied in more detail:
\begin{enumerate}
\item Multi-scale problems;
\item Colour-annihilation form factors;
\item Top quark mass schemes.
\end{enumerate}
These are discussed in \secsRef{sec:multiscale}, \ref{sec:II}, and \ref{sec:masses}, respectively. 

In \secRef{sec:nonGauss}, we give a brief summary of a recent proposal for a frequentist statistical treatment of ``errors on errors''~\cite{Cowan:2018lhq,Canonero:2023sua,Canonero:2025vxq}, which exhibits non-Gaussian properties. This may be of interest for theory uncertainties and systematic uncertainties more broadly~\cite{Canonero:2024kzk}. Finally, we summarise and give an outlook in \secRef{sec:summary}.



\section{Effects that go beyond 7-point variations}
\label{sec:beyond}
For a Born-level matrix element proportional to $\alpha_s^n$, with a central renormalisation-scale choice of $\mu_0$, changing the renormalisation-scale prefactor to $\mu' = k\mu_0$ generates extra terms 
\begin{equation}
\big|M(\mu')\big|^2 ~=~
\Big(1\,-\,\alpha_s(\mu_0) b_0 \ln(k^2)\,+\,{\cal O}(\alpha_s^2)\Big)^n \,\big|M(\mu_0)\big|^2\,,\label{eq:muRvar}
\end{equation}
with $b_0 = (33-2n_f)/(12\pi) \sim {\cal O}(1)$. 
Similarly, to estimate the uncertainty associated with the choice of factorisation scale\footnote{I.e., the scale at which the collinear singularities from initial-state emissions are split between the PDFs and the hard cross section in collinear factorisation.}, one performs the same change as in \eqRef{eq:muRvar} in the DGLAP anomalous dimensions. 
Loop-level matrix elements contain explicit renormalisation- and factorisation-scale-dependent terms that cancel against those in \eqRef{eq:muRvar} up to the given loop order, pushing the remaining variation with $k$ to higher orders. 
At the completely pragmatic level, renormalisation-and factorisation-scale variations are thus simply a method to inject explicit higher-order terms, which are generically of order unity times powers of $\alpha_s$ and which can be cancelled in a well-defined way up to a given order. It is also convenient: if a calculation has been done in terms of arbitrary renormalisation and factorisation scales, little additional work is needed to vary these scales. And, besides discussions about how to choose the central scale (which, admittedly, can be fraught), it is also quite universal: the only aspects of \eqRef{eq:muRvar} that are process-specific are the central scale choice, $\mu_R$, and the power, $n$. This attractive combination, of convenience, universality, and systematic cancellation through a given order, has made renormalisation- and factorisation-scale variations the dominant paradigm for perturbative uncertainty estimates for decades.

However, \eqRef{eq:muRvar} also illustrates many of the issues that can make scale variations ``false friends'' in the context of estimating perturbative uncertainties:
\begin{itemize}
\item The Born-level matrix elements for processes such as diphoton production, weak boson fusion, or (multiple) vector-boson production do not involve any powers of the strong coupling. For such processes, $n=0$ and hence  \eqRef{eq:muRvar} generates zero variation at leading order. Obviously, this does not imply that the perturbative uncertainty is zero, only that renormalisation-scale variation cannot access it.
\item The higher-order terms generated by \eqRef{eq:muRvar} are directly proportional to the lower-order matrix elements. This does not capture genuinely new structures that can appear at higher orders, including new kinematic regions, new helicity structures, new colour channels, and interference with other Born-level processes. The question of new flavour structures is also sometimes raised in this context, though one should take note that PDFs do include inclusive sums (in the collinear approximation) over such channels as are accessible via PDF evolution.
\item For multiscale problems, the notion of a single characteristic central scale, $\mu_R$, that one can make order-unity variations around, is inherently problematic. Instead, one should at the very least vary multiple scales. This issue is becoming ever more relevant as the complexity (i.e., final-state multiplicity) of the processes we study grows. We return to this in \secRef{sec:multiscale}. 
\item Related to the above, multiscale hierarchies with potentially dramatic impacts on perturbative calculations can appear in very many contexts, including:
\begin{itemize}
\item Differential observables often span a large range of underlying kinematics, extreme regions of which will typically be associated with  logarithms of a small parameter. Typical examples include transverse-momentum distributions of Higgs, $\mathrm{W}$, or $\mathrm{Z}$ bosons at hadron colliders, or multijet production at asymptotically high energies~\cite{Andersen:2026oge}. Indeed, the leading power of such logarithmic contributions is known to increase with perturbative order both in the soft (Sudakov) and high-energy (BFKL) limits. This is not captured by scale variation, which thus inevitably underestimates uncertainties in regions where logarithmic resummation is necessary.
\item Even at the level of integrated cross sections, large scale ratios can appear, for instance for production of off-shell gauge bosons with masses much lower than the collision energy, or, on the contrary, very high masses close to the threshold region.
\item Real-world observables are typically not fully inclusive in the Born-level kinematics but involve cuts or vetoes that reflect experimental fiducial regions, background suppression, and/or the specific characteristics of a signal that one wishes to study. Tight cuts or other tight restrictions (such as a very small jet radius, $\Delta R$~\cite{Generet:2025vth}) effectively generate observable-induced multi-scale hierarchies analogous to those for extreme values of differential observables (i.e., typically involving logarithms of a small parameter).
\end{itemize}
\item In processes that involve annihilating colours, such as $gg\to H$ or $q\bar{q}\to V$, terms $\propto -\alpha_s\ln^2(-1) = \alpha_s \pi^2$ can generate numerically large higher-order corrections. We return to this in \secRef{sec:II}. Note also that there is substantial overlap between this set of processes and those mentioned under the first point above which have $n=0$.
\item For processes involving massive quarks (including when they only appear in loops), higher-order calculations require a precise definition of the renormalised quark mass, via a mass scheme. Simplistic mass scheme variations can risk inducing artificial terms that systematically \emph{overestimate} the true uncertainty. We return to this in \secRef{sec:masses}.
\item Finally, we should also briefly mention QED, for which higher-order corrections are important for precision-frontier experiments such as MUonE or Belle II, and for FCC-ee studies. Especially at lower energies near $\sqrt{s}\lesssim$ 10 GeV, calculations are usually performed in the on-shell scheme for $\alpha_{\rm em}$. This means that the physical result has no scale dependence at all and therefore scale variation is not possible. Uncertainties are usually estimated by multiplying the known order by $\alpha_{\rm em} L/\pi$ using whatever large logarithm $L$ seems most appropriate. However, this is much worse than 7-point variation is for QCD calculations and, the true truncation error is only known once the next order has been calculated.
\end{itemize}
Common to all of these is that they are not accessed by conventional renormalisation- or factorisation-scale variations. Nevertheless,  it is often possible to predict their presence (or absence) on general grounds. It therefore makes sense to consider whether such qualitative predictions can, with some degree of ease (or not), be turned into quantitative uncertainty estimates.  

There are of course many other sources of perturbative uncertainties as well, which are, likewise, not accessed by scale variations. These include finite-width effects, power corrections, and factorisation-breaking effects, to name but a few. We do not claim our explorations here are exhaustive but consider it useful to summarise the outcome of some of the discussions that took place in connection with the Aspen workshop.

We note that, in addition to resummations of the standard kinds of soft and collinear logarithms, there has been important recent progress on understanding (and resumming) so-called super-leading logarithms~\cite{Becher:2021zkk,Becher:2023mtx,Becher:2024kmk,Becher:2024nqc}. These originate from coherence-violating initial-state effects and could hence also be called coherence-violating logarithms (CVLs). They appear at quite high order in the perturbative series, starting from ${\cal{O}}(\alpha_s^3)$, and not at leading colour so can safely be ignored for uncertainty estimates at LO and presumably also at NLO except possibly for cases of extremely aggressive jet vetos for which they can reach ${\cal O}(10\%)$~\cite{Becher:2024nqc}. Starting from NNLO, we would recommend to allow for an uncertainty due to missing CVLs at the next order if their presence cannot be ruled out for the process/problem at hand.



\subsection{Multiscale problems}
\label{sec:multiscale}

In complex processes, such as $\mathrm{Z}/\mathrm{W}$ + multiple jets, $\t\bar{\t}$ + jets, or multi-boson production (+ jets), renormalisation generally induces higher-order terms that involve logarithms of many different physical scales. For genuine multiscale problems (with more than a single relevant scale ratio), no single scale choice captures all of these terms --- nor can the full range of scale dependence be explored by a simple factor-2 variation around such a choice.

\begin{figure}[tp]
\centering
\includegraphics*[width=0.99\textwidth]{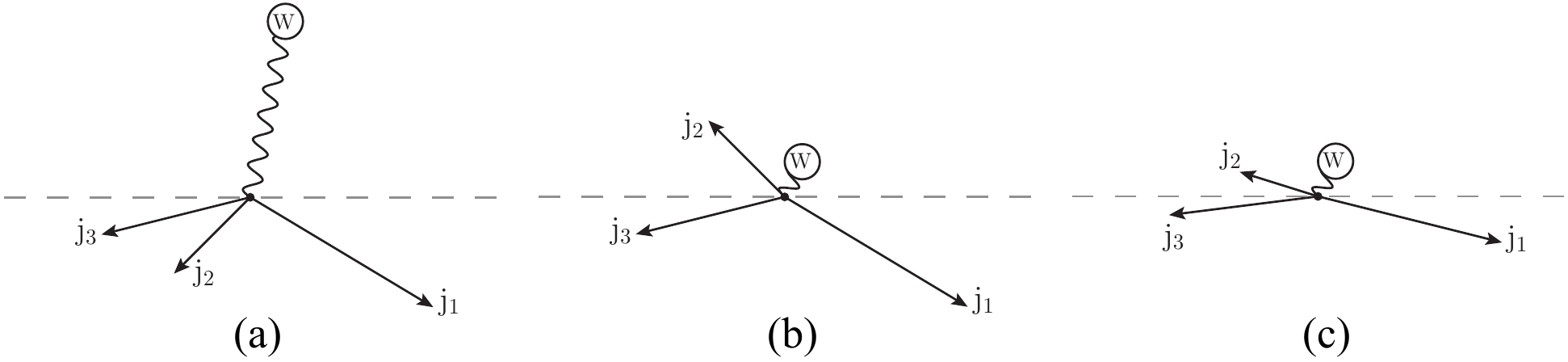}
\caption{Illustration of the three kinematic configurations for $\W$ + 3 jets discussed in the text.\label{fig:W3abc}}
\end{figure}

\begin{figure}[tp]
\centering
\includegraphics*[width=0.33\textwidth]{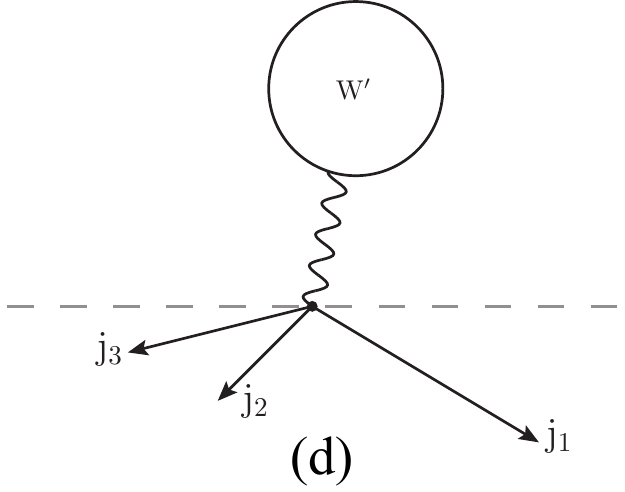}
\caption{Illustration of a heavy $\W'$  balancing against 3 jets.\label{fig:W3d}}
\end{figure}

A particularly well-studied example of a process that offers a broad range of multi-scale aspects is $\mathrm{W}$ + 3 jets~\cite{Bauer:2009km,Berger:2009ep,Melnikov:2009wh,Hoche:2012tae,Andersen:2012gk,Anger:2017nkq}. In the original NLO study~\cite{Berger:2009ep}, two examples of multi-scale kinematic regions were highlighted:
\begin{enumerate}[label=\alph*)]
\item A high-$p_T$ $\W$ boson balancing against a system of jets, illustrated in \figRef{fig:W3abc}a. 
\item A lower-$p_T$ $\W$ boson accompanied by a system of hard jets, illustrated in \figRef{fig:W3abc}b.
\end{enumerate}
In addition, we also consider two more cases which are representative of kinematic regions dominated by bremsstrahlung,
\begin{enumerate}[label=\alph*), start=3]
\item The same $p_{T \W}$ as in (b), but accompanied by a system of softer jets, making $M_\W$ the largest single-particle scale in the problem. Illustrated in \figRef{fig:W3abc}c.
\item The same jets as in (a), but with $m_\W$ increased by a factor of 10 (e.g., representative of a hard new-physics process), making $m_{\W'}$ the largest single-particle scale in the problem. Illustrated in \figRef{fig:W3d}.
\end{enumerate}
Specifically, we consider the mass and $p_T$ values given in \tabRef{tab:W3}. 
\begin{table}[t]
\centering
\begin{tabular}{clrrrrc}
\toprule
& & (a) & (b) & (c) & (d) &\\
\midrule 
&$m_\mathrm{W}$~ & 80 & 80 & 80 & 800 &\\
&$p_{T \mathrm{W}}$~ & ~~500 & ~~~50 & ~~50 & ~~~500 & \\
\midrule 
&$p_{T \mathrm{j}_{\mbox{\scriptsize 1}}}$ & 300 & 300 & 60 & 300 &\\
&$p_{T \mathrm{j}_{\mbox{\scriptsize 2}}}$ & 200 & 200 & 40 & 200 & \\
&$p_{T \mathrm{j}_{\mbox{\scriptsize 3}}}$ & 100 & 100 & 20 & 100 & \\
\bottomrule
\end{tabular}
\caption{Single-particle scales (in GeV) used for the $\W+3$ jets scale-setting and scale-variation examples in this work.
\label{tab:W3}}
\end{table}
In each scenario, several different choices of renormalisation scales have been proposed and used. We will consider the following:
\begin{enumerate}
\item The scalar sum of transverse masses,
\begin{equation}
H_T = \sqrt{m_\mathrm{W}^2 + p_{T \mathrm{W}}^2} ~+~ \sum_{j=1}^3 |p_{T\mathrm{j}}|\,,
\end{equation}
\item The scalar sum of jet $p_T$ values, 
\begin{equation}
H_{T}^\mathrm{jets} = \sum_{j=1}^3 |p_{T \mathrm{j}}|\,,
\end{equation}
\item The largest transverse mass, 
  \begin{equation} 
 m_T^\mathrm{max} = \max\Big( m_{T \mathrm{W}},\, p_{T \mathrm{j}_{\mbox{\scriptsize 1}}}\Big)\,,
  \end{equation}
\item The transverse momentum of the leading jet,
  \begin{equation} 
  p_{T\mathrm{j}_{\mbox{\scriptsize 1}}}\,,
  \end{equation}
\item The geometric mean of transverse masses, 
\begin{equation}
\overline{m}_T = \Big(m_{T\mathrm{W}}\,
p_{T \mathrm{j}_{\mbox{\scriptsize 1}}}\, 
p_{T \mathrm{j}_{\mbox{\scriptsize 2}}}\, 
p_{T \mathrm{j}_{\mbox{\scriptsize 3}}}
\Big)^{1/4}\,, \label{eq:geoAll}
\end{equation}
\item The geometric mean of jet $p_\perp$ values,
\begin{equation}
\overline{p}^\mathrm{jets}_T = \Big(
p_{T \mathrm{j}_{\mbox{\scriptsize 1}}}\, 
p_{T \mathrm{j}_{\mbox{\scriptsize 2}}}\, 
p_{T\mathrm{j}_{\mbox{\scriptsize 3}}}
\Big)^{1/3}\,, \label{eq:geoJets}
\end{equation}
\item The transverse mass of the $\mathrm{W}$ boson, 
\begin{equation}
m_{T\mathrm{W}} = \sqrt{m_\mathrm{W}^2 + p_{T \mathrm{W}}^2}\,.  
\end{equation}
\end{enumerate}
Note that it was remarked already in \cite{Berger:2009ep} that at least the last one, $m_{T \mathrm{W}}$, is disfavoured for configurations like (b) where a soft $\W$ is accompanied by much harder jets. We include it mostly for comparison and exclude it from our consideration of reasonable uncertainty ranges. We also note that the transverse mass of the $\mathrm{W}$ boson used here will be slightly larger than the scalar sum of the lepton + neutrino $p_T$ values commonly used when defining $H_T$. 

Other definitions, e.g., involving the invariant mass of the hadronic system~\cite{Berger:2009ep}, could also be included but would require more fine-grained specification of the kinematics than we intend here. 

The geometric means used in \eqsRef{eq:geoAll} and \eqref{eq:geoJets} are motivated by the fact that this is the simplest brute-force way to turn a single-logarithmic multiscale problem into a single-scale one,
\begin{eqnarray}
\alpha_s(\mu_1)\alpha_s(\mu_2)\cdots\alpha_s(\mu_n) & = &
\prod_{i=1}^{n} \alpha_s(\mu) \left(1 +
b_0\,\alpha_s\ln\left(\frac{\mu^2}{\mu_i^2}\right) + {\cal O}(\alpha_s^2)\right)
\nonumber\\[2mm]
& = & \alpha_s^n(\mu) \left(1 + b_0\, \alpha_s \ln \left(
 \frac{\mu^{2n}}{\mu_1^2\mu_2^2\cdots\mu_n^2}\right) +  {\cal
   O}(\alpha_s^2) \right)\\[2mm]
 & = & \alpha_s^n(\overline{\mu}) \Big(1 +  {\cal
   O}(\alpha_s^2) \Big)
  ~,
\end{eqnarray}
with $\overline{\mu}$ the geometric mean of the individual scales, $\overline{\mu}^n = \prod_{i=1}^n \mu_i$.

Alternatively, CKKWL-type~\cite{Catani:2001cc,Lonnblad:2001iq} sequential jet-clustering scales could be used. This would tend to increase the $p_T$ scales of the harder jets, and since we already include $H_T$ as a reasonable maximum scale, we elect to use the unclustered jet $p_T$ scales as reasonable minimum ones. This reasoning is of course specific to our aim of asking not whether this or that choice of central scale is most appropriate but whether we can place reasonable \emph{bounds} on the renormalisation-induced terms, from above and from below. 

\begin{figure}[tp]
\centering
\hspace*{-3mm}\includegraphics*[width=0.492\textwidth]{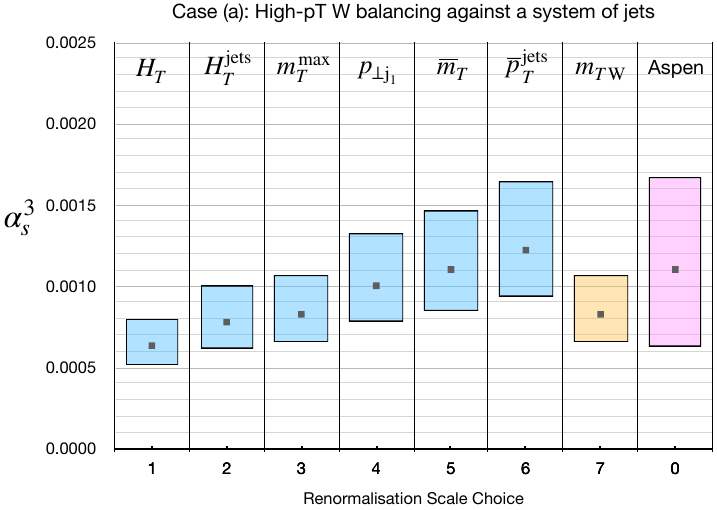}\hfill
\includegraphics*[width=0.492\textwidth]{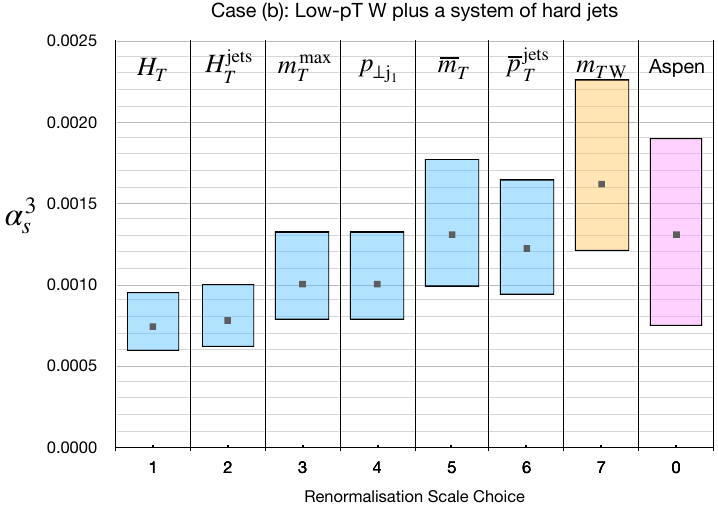}\hspace*{-3mm}\\[8mm]
\hspace*{-3mm}\includegraphics*[width=0.492\textwidth]{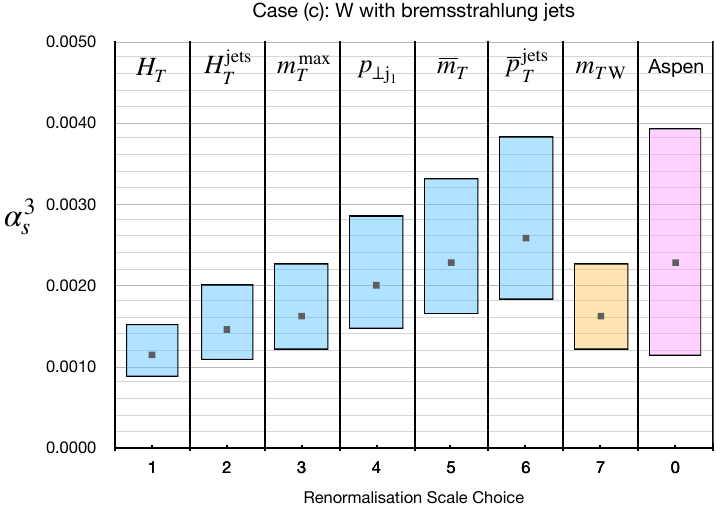}\hfill
\includegraphics*[width=0.492\textwidth]{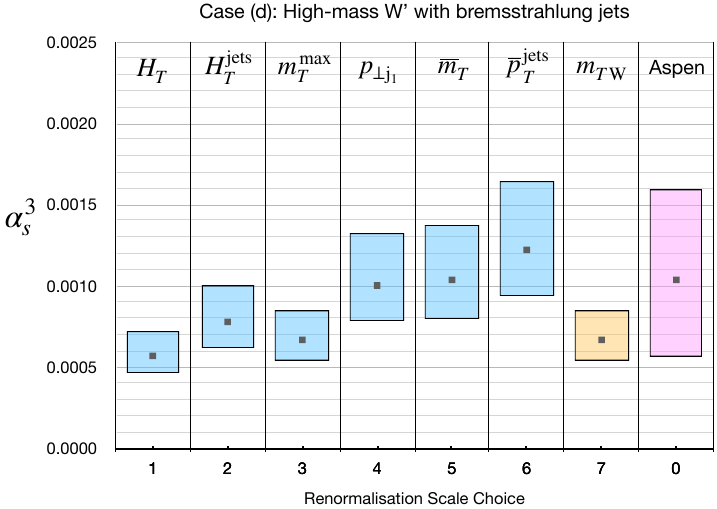}\hspace*{-3mm}
\caption{Illustration of the effect of renormalisation-scale variations on $\alpha_s^3$, 
for four representative cases of $W+3\,\mathrm{jets}$ kinematics (a, b, c, and d). We compare seven different choices of central scale with conventional factor-2 variations (blue and orange shaded bands) to the range spanned by our tentative prescription for a conservative multiscale variation (last column, pink shaded bands). See text for details.\label{fig:muVarTables}}
\end{figure}

In \figRef{fig:muVarTables}, we illustrate the sizes of $\alpha_s^3(\mu)$ for each of the renormalisation-scale choices above (square dots), along with bands showing the variation generated by factor-2 variations around those choices, for each of the kinematic scenarios (a) -- (d), using a 2-loop running 5-flavour $\alpha_s$ with $\alpha_s(m_Z) = 0.118$. 
The $m_{T\W}$ choice which is disfavoured for case (b) is set apart by orange shading, and in the last column we illustrate with pink shading a possible choice of conservative multiscale variation, elaborated on below, with, for these examples, the highest scale (lowest $\alpha_s$ value) defined by $H_T$, the lowest scale (highest $\alpha_s$ value) defined by the geometric mean of $p_{T3}$ and the (geometric) mean of the other scales in $H_T$, divided by 2, and the central dot at the value obtained for $\mu = \overline{m}_T$. 

These bands obviously only illustrate the uncompensated scale variations, i.e., the re\-nor\-mal\-i\-sation-scale uncertainty that one would attach to an LO calculation of $\W$ + 3 jets with the corresponding scale choices. At NLO and beyond, these variations would be partly compensated by the explicit $\mu$-dependent logarithms in the matrix elements. In such contexts, it is important to note that the inflexion of the scale dependence means that one usually has to probe several scales inside the given range of $\mu$ values since the maximum/minimum cross section may not be at the extremal points. 

A simple general recipe that could be considered conservative within the limited context of scale variations, illustrated by the last (pink shaded) column in \figRef{fig:muVarTables}, is to let the maximum scale be defined by:
\begin{equation}
 \mu_\mathrm{\uparrow}~=~H_T~=~\sum_i |m_{Ti}|\,,
\end{equation}
where the sum over $i$ generically includes any particles that are produced in vertices that involve coloured lines,  $m_{Ti}=\sqrt{m_i^2 +p_{Ti}^2}$ for massive particles, and $m_{Ti}=p_{Ti}$ for massless QCD partons. For processes which feature a hard vertex that does not involve any QCD partons at all, such as VBF Higgs production, the Higgs boson itself would be left out of the sum over $i$, while the forward jet $p_{Ti}$ values could be replaced by approximate $t$-channel momentum transfers of order $\sqrt{p_{Ti}^2+m_W^2}$.

For a central scale choice, 
\begin{equation}
\mu_{0}~=~\overline{m}_T~=\bigg(\prod_{i} m_{Ti}\bigg)^{1/n}\,.
\end{equation}
represents the geometric mean of the individual terms entering in the $H_T$ sum.

And for a minimum scale choice, one option is to let the smallest scale in the problem have a relatively higher weight in the geometric mean, 
\begin{equation}
 \mu_{\downarrow}~=~\frac{\sqrt{ \,\bar{\mu}_{i\neq \mathrm{min}}\,\min(m_{Ti})}}{2}\,,
\end{equation}
i.e., the geometric mean of the smallest term entering in the $H_T$ sum and the (geometric) mean of all of the other terms in $H_T$, divided by two.

\begin{figure}[t]
\centering
\includegraphics[width=0.75\linewidth]{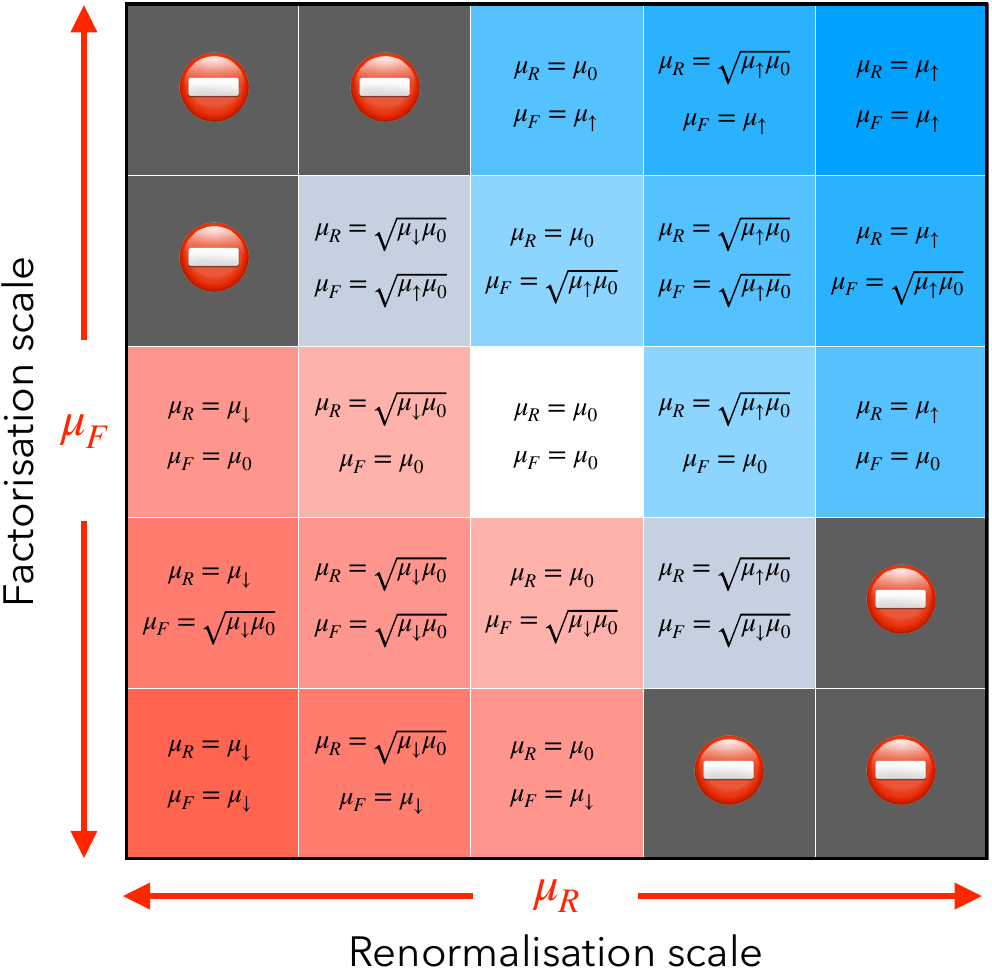}
\caption{Illustration of 19-point scale variations. Compared to 7-point variations, two additional scales are added, at the geometric midpoints between the central and extremal variations. 
\label{fig:19point}
}
\end{figure}
In \figRef{fig:19point}, we show a 19-point generalisation of the 7-point variations, with two additional intermediate scales at the (geometric) midpoints between the central and extremal scale choices intended to give a more complete exploration of the scale-uncertainty range for processes computed at higher orders, for which the scale dependence may not be monotonic. (If a more complete exploration is desired, one may of course decide to add more intermediate points.)

\subsection{Colour-annihilation form factors}
\label{sec:II}

Scattering processes that involve colour-annihilation channels (a.k.a.\ initial-initial colour flows) are known to be associated with form factors of the generic form~\cite{Radyushkin:1982kg,Magnea:1990zb,Ahrens:2008qu}
\begin{equation}
\exp\left[\frac{-\alpha_s}{2\pi}  {\cal C}\ln^2(-1) \right]
= \exp\left[\alpha_s\frac{{\cal C} \pi }{2}\right]\,,
\end{equation}
where the colour factor ${\cal C} = C_A = 3$ for gluon-gluon fusion in the colour-singlet channel~\cite{Ahrens:2008qu,Ahrens:2009cxz} while  ${\cal C}=C_F = 4/3$ for colour-singlet quark-antiquark annihilation~\cite{Magnea:1990zb,Ahrens:2008qu,Becher:2007ty}. 

Physically, these form factors are thought to originate from non-cancellations between  diagrams such as those shown in \figRef{fig:diagramsFFII}:
\begin{figure}[tp]
\centering
\includegraphics*[width=0.49\textwidth]{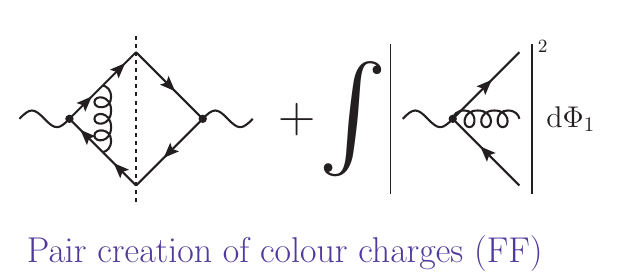}\hspace*{-2mm}%
\raisebox{17.5mm}{\LARGE,}\hfill\includegraphics*[width=0.49\textwidth]{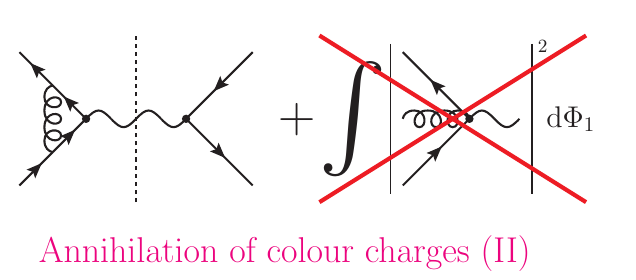}\hspace*{-3mm}
\caption{\label{fig:diagramsFFII}NLO QCD corrections to (left) hadronic \Z decay --- exemplifying a Born-level final-final (FF) colour flow --- and (right) the analogous fully crossed diagrams for Drell-Yan production --- exemplifying a Born-level initial-initial (II) colour flow.}
\end{figure}
\begin{itemize}
\item For pair creation of final-state colour charges (left-hand pane of \figRef{fig:diagramsFFII}, a.k.a.\ final-final colour flows), cancellations between the real and virtual corrections leave a rational remainder\footnote{When the emission phase space is integrated over inclusively, without cuts, and omitting logarithms induced by renormalisation.}. For the case illustrated here --- NLO $\Z$ decay --- this just produces the well-known factor $\mathrm{d}\Gamma^{\mathrm{NLO}} = (1+\alpha_s/\pi)\mathrm{d}\Gamma^{\mathrm{LO}}$, which obeys the expected ${\cal O}(\alpha_s)$ parametric suppression.
\item For annihilation of initial-state colour charges (right-hand pane of \figRef{fig:diagramsFFII}, a.k.a.\ initial-initial colour flows), the sum over real corrections does not include the crossed-out diagram (``emission into the initial state''). This leaves the uncancelled $-\ln^2(-1) = -(i\pi)^2 = \pi^2$ in the squared amplitude, cast in exponentiated (all-orders) form in the formula above.
\item Analogous factors do not appear for DIS- or VBF-type (initial-final) colour flows, since the change of one of the invariants from spacelike to timelike flips the sign of the $-1$ in the logarithm, making it vanish.
\end{itemize}

This suggests a simple general recipe that can be easily implemented in event generators. In all current general-purpose MC event generators, each event is associated with an LC colour flow. This is true even if the weight for the corresponding parton configuration is computed at full colour, and is then done by stochastic selection according to the relative weights of each LC flow for the given phase-space point.

\begin{figure}[tp]
\centering
\includegraphics*[scale=0.25]{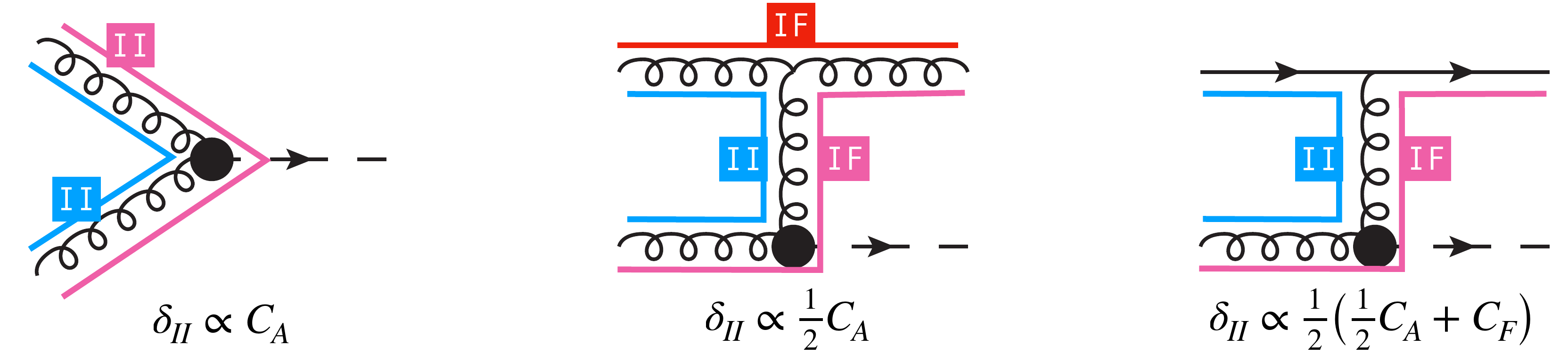}
\caption{
LC colour flows for $gg\to H$ and $gg\to H+\mathrm{jet}$, with indications of initial-initial (II) and initial-final (IF) colour-dipole connections, and corresponding proposed colour factors for LO II form-factor uncertainties. 
\label{fig:colFlowH}}
\end{figure}
\begin{figure}[tp]
\centering
\includegraphics*[scale=0.25]{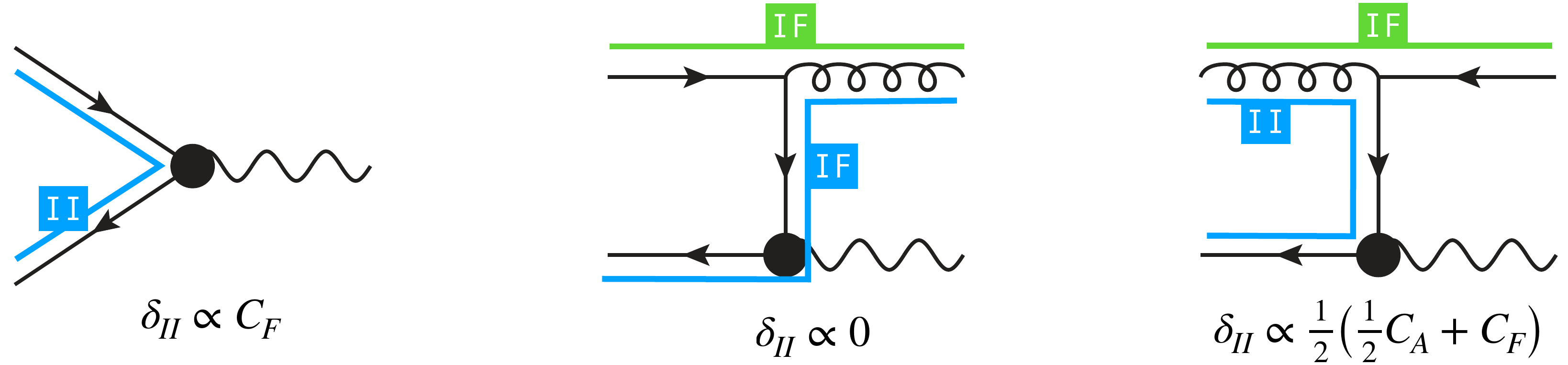}
\caption{
LC colour flows for $q\bar{q}\to Z$ and $q\bar{q}\to Z+\mathrm{jet}$, with indications of initial-initial (II) and initial-final (IF) colour-dipole connections, and corresponding proposed colour factors for LO II form-factor uncertainties.
\label{fig:colFlowZ}}
\end{figure}
\begin{figure}[tp]
\centering
\includegraphics*[scale=0.25]{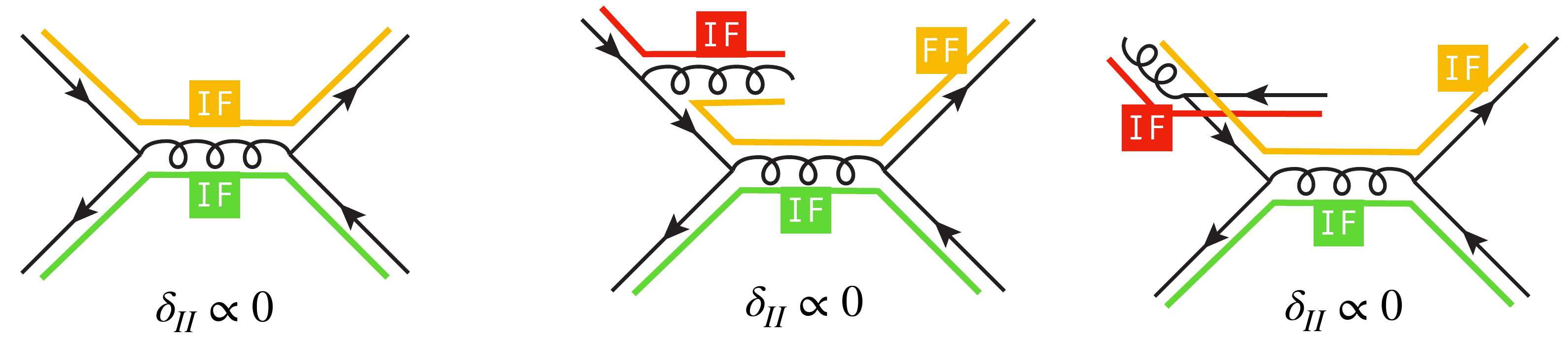}
\includegraphics*[scale=0.25]{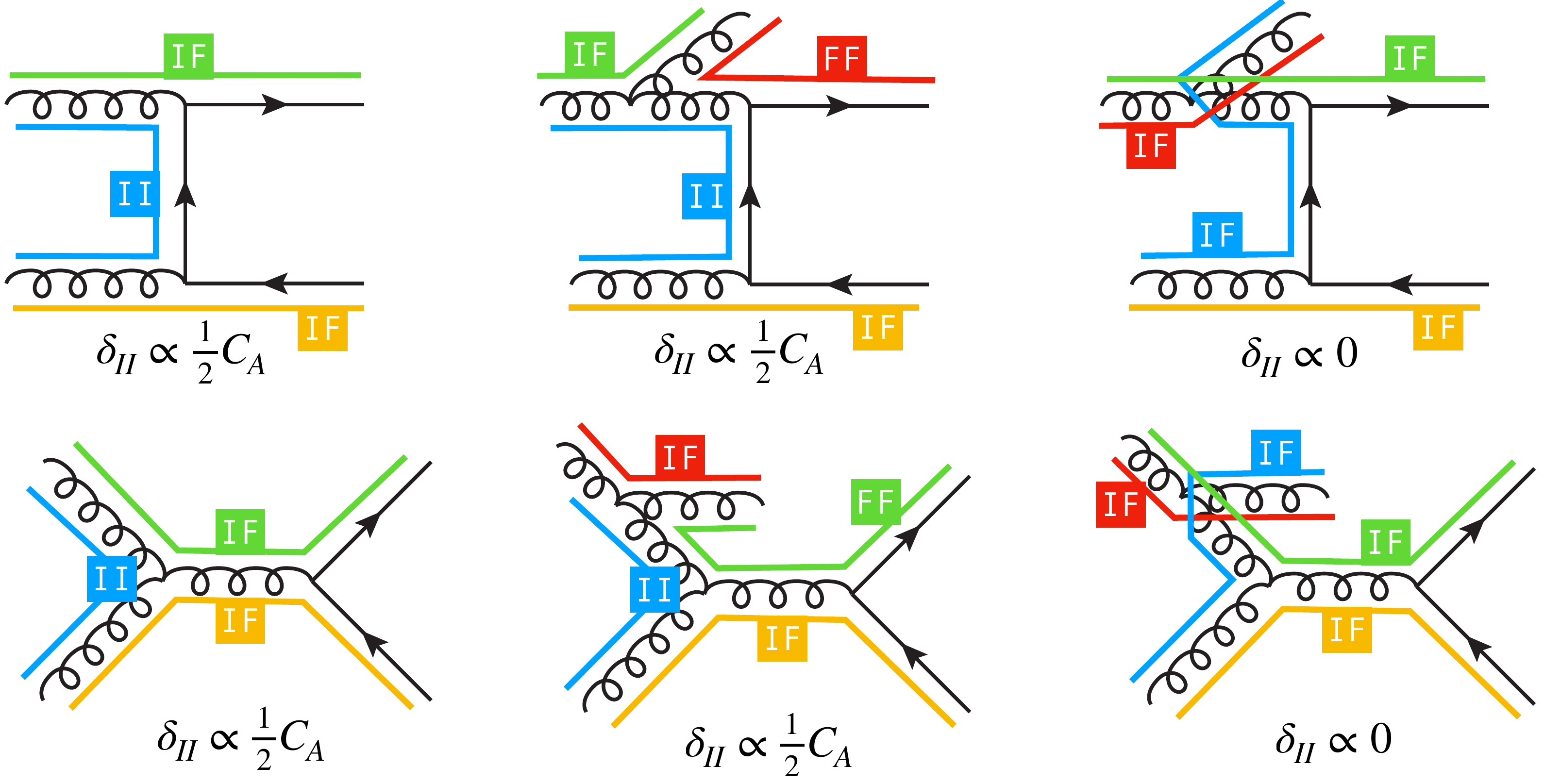}
\caption{
LC colour flows for $q\bar{q}\to t\bar{t}$, $q\bar{q}\to t\bar{t}+\mathrm{jet}$ (ISR), $gg\to t\bar{t}$, and $gg\to t\bar{t}+g$ (ISR), with indications of initial-initial (II), initial-final (IF), and final-final (FF) colour-dipole connections, and corresponding proposed colour factors for LO II form-factor uncertainties. 
\label{fig:colFlowTT}}
\end{figure}
Figs.~\ref{fig:colFlowH} -- \ref{fig:colFlowTT} illustrate a number of examples of such flows, for would-be MC events for $gg\to H$ and $H+\mathrm{jet}$ (\figRef{fig:colFlowH}), $q\bar{q}\to Z$ and $Z+\mathrm{jet}$ (\figRef{fig:colFlowZ}), 
$q\bar{q}\to t\bar{t}$ and $t\bar{t}+\mathrm{jet}$ (\figRef{fig:colFlowTT}, top row, ISR only), and $gg\to t\bar{t}$ and $t\bar{t}+g$ (\figRef{fig:colFlowTT}, bottom two rows, ISR only).

In each MC event, we can thus count the number of (LC) initial-initial colour dipoles in the event. For each such dipole, we propose to associate an initial-initial form-factor exponent proportional to $C_A/4$ or $C_F/2$, respectively, for each gluon or (anti)quark dipole end. Thus, e.g., for gluon-gluon fusion in a colour-singlet channel, the total of four initial-initial octet dipole ends (leftmost in \figRef{fig:colFlowH}) sum to the correct ${\cal C}=C_A$, and similarly the two II triplet dipole ends for quark-antiquark annihilation in a colour-singlet channel sum to ${\cal C}=C_F$ (leftmost in \figRef{fig:colFlowZ}). For gluon-gluon fusion in a colour-octet channel (leftmost diagrams in the bottom two rows of \figRef{fig:colFlowTT}), we have two initial-initial octet dipole ends which sum to ${\cal C}=C_A/2$, while  for, say, quark-gluon scattering in a colour-triplet channel (rightmost diagrams in \figsRef{fig:colFlowH} and \ref{fig:colFlowZ}), our effective II colour factor becomes ${\cal C}=(C_A+2C_F)/4$. 

To obtain a rough estimate of the size that may be expected for the initial-initial form factor for a generic process, we thus propose the
following prescription at LO: 
\begin{equation}
\delta_{II}^\mathrm{LO} =  \frac{\pi \alpha_s(\mu_F^2)}{2}\left(C_A
f_{gg1} + \frac12C_A f_{gg8} +
\frac{C_A+2C_F}{4} f_{qg3} + C_F f_{q\bar{q}1}\right) \,,\label{eq:deltaII}
\end{equation}
where $f_{ijm}$ denotes the fraction of the LO cross section that
corresponds to $ij$ initial states in a colour-$m$
channel\footnote{For more exotic colour topologies, like incoming $gg$
pairs in decuplet representations, incoming $qg$ pairs in sextet
representations, or incoming $qq$ pairs in antitriplet
representations, additional coefficients would need to be added. 
We leave that for a possible future extension.}.
At higher orders, the corresponding estimates are:
\begin{eqnarray}
\mathrm{NLO} & : & \frac{\pi^2 \alpha_s^2(\mu_F^2)}{8} \left( C_A^2
  f_{gg1} + \frac14C_A^2 f_{gg8} + \left(\frac{C_A+2C_F}{4}\right)^2
  f_{qg3} + C_F^2 f_{q\bar{q}1}  \right)\,, \\[2mm]
\mathrm{NNLO} & : & \frac{C_A^3 \pi^3 \alpha_s^3(\mu_F^2)}{48}\left( C_A^3
  f_{gg1} + \frac18C_A^3 f_{gg8} + \left(\frac{C_A+2C_F}{4}\right)^3
  f_{qg3} + C_F^3 f_{q\bar{q}1}  \right)\,, \\[2mm]
\mathrm{N3LO} & : & \frac{C_A^4 \pi^4 \alpha_s^4(\mu_F^2)}{384}\left( C_A^4
  f_{gg1} + \frac1{16}C_A^4 f_{gg8} + \left(\frac{C_A+2C_F}{4}\right)^4
  f_{qg3} + C_F^4 f_{q\bar{q}1}  \right)\,,
\end{eqnarray}
and the $f_{ijm}$ fractions should presumably likewise be estimated from the relevant N$^n$LO cross sections. 

In the context of MC event generators, the randomly sampled
event-by-event partonic scattering channels and colour flows can be used to estimate the fractions $f_{ijm}$ directly, e.g., based on the
flavour and Les Houches (LH) colour-tag information~\cite{Boos:2001cv}
of the incoming
partons in partonic event records and LHE
files~\cite{Alwall:2006yp,Andersen:2014efa}. The uncertainty  
due to the next-order II form factor can thus be accounted for by including a multiplicative reweighting factor, $1+\delta_{II}$, for each event given by numerical coefficients listed in
tab.~\ref{tab:diiCol}. In tab.~\ref{tab:deltaII-sh} we give numerical values for the uncertainties, for a selection of representative LHC processes at $\sqrt{s}\,=\,14\,\mathrm{TeV}$, with the $f_{ijk}$ calculated with \sherpa \cite{Sherpa:2024mfk} using \comix \cite{Gleisberg:2008fv}, with $\alpha_s(m_Z)=0.118$, 2-loop running with 5 flavours, the NNPDF~3.0 NNLO PDF set~\cite{NNPDF:2014otw}, and summing inclusively over all branching fractions of the respective resonances.
\begin{table}[tp]
\centering
\begin{tabular}{lcccc}
  \toprule
$\delta_{II}$  & LO & NLO & NNLO & N3LO \\
  \midrule
  $g=g$ ($gg$ singlet) &
  $\frac{C_A\pi}{2}\alpha_s \sim 4.7\alpha_s$ &
  $\frac{C_A^2\pi^2}{8}\alpha_s^2 \sim 11\alpha_s^2$ &
  $\frac{C_A^3\pi^3}{48}\alpha_s^3 \sim 17\alpha_s^3$ &
  $\frac{C_A^4\pi^4}{384}\alpha_s^4 \sim 21\alpha_s^4$   
  \\[2mm]
  $g-g$ ($gg$ octet) &
  $\frac{C_A\pi}{4}\alpha_s \sim 2.4\alpha_s$ & 
  $\frac{C_A^2\pi^2}{32}\alpha_s^2 \sim 2.8\alpha_s^2$ & 
  $\frac{C_A^3\pi^3}{384}\alpha_s^3 \sim 2.2\alpha_s^3$ & 
  $\frac{C_A^4\pi^4}{6144}\alpha_s^4 \sim 1.3\alpha_s^4$ \\[2mm]
  $q-g$ ($qg$ triplet) &
  $\sim 2.2\alpha_s$ &
  $\sim 2.5\alpha_s^2$ &
  $\sim 1.8\alpha_s^3$ &
  $\sim 1.0\alpha_s^4$ 
  \\[2mm]
  $q-\bar{q}$ ($q\bar{q}$ singlet) & 
  $\frac{C_F\pi}{2}\alpha_s \sim 2.1 \alpha_s$ &
  $\frac{C_F^2\pi^2}{8}\alpha_s^2 \sim 2.2 \alpha_s^2$ & 
  $\frac{C_F^3\pi^3}{48}\alpha_s^3 \sim 1.5 \alpha_s^3$ & 
  $\frac{C_F^4\pi^4}{384}\alpha_s^4 \sim 0.8 \alpha_s^4$ 
  \\[1mm]\bottomrule
\end{tabular}
\caption{Estimated single-sided theory uncertainty due to missing II form-factor
  terms at the next order, ordered by colour channel.
\label{tab:diiCol}}
\end{table}
One sees clearly the very large enhancement for gluon-gluon
annihilation in the colour-singlet channel, and that it continues deep
into the perturbative series. The coefficients for the other
colour-annihilation channels are more modest, especially beyond LO.


\begin{table}[tp]
\centering
\begin{tabular}{p{2cm}cccccccc}
\toprule 
$\delta_{II}$ & ggH & Z & ZZ &  {Z+j$_{30}$} & Z+j$_{100}$ & $t\bar{t}$ &
jj$_{50}$ & jj$_{200}$ 
 \\ 
\midrule 
LO   & $+59\%$ & $+27.6\%$ & $+24.7\%$ & $ +21.7\%$ & $+21.5\%$ & $+22.1\%$ & $+13.4\%$ & $+10.1\%$ \\
NLO$_\text{approx.}$ & $+17\%$ & $+3.8\%$ & $+3.1\%$ & $+3.1 \%$ & $+2.7\%$ & $+2.8\%$ & $+2.0\%$ & $+1.2\%$ \\
NLO  & $+18\%$ & $+3.9\%$ & $+3.1\%$ & $+3.3 \%$ & $+2.4\%$ & $+3.0\%$ & $+1.8\%$ & $+1.2\%$ \\
\bottomrule
\end{tabular}  
\caption{Examples of single-sided initial-initial form-factor uncertainty
 estimates obtained with \sherpa/\comix, for a selection of hard processes in $pp$ collisions at 14 TeV CM
 energy. The arguments used to evaluate $\alpha_s$ in each case are,
 respectively,  $m_H/2$, $m_Z/2$, $m_Z$, 54 GeV, 120 GeV, $m_t$, 50 GeV, and
 200 GeV, using $\alpha_s(m_Z) = 0.118$ and 2-loop running. NLO$_\text{approx.}$ corresponds
 multiplying the LO $f_{ijk}$ with NLO factors, while in the last line they are evaluated at NLO. Note that the LO estimate for $\mathrm{Z}+\mathrm{j}_{30}$ was cross checked using the LO LC $qg$ fraction computed with \nnlojet, which gave the same result.
}
\label{tab:deltaII-sh}
\end{table}

\begin{table}[tp]
\centering
\begin{tabular}{p{1.5cm}ccccc}
\toprule 
Order & ggH & Z & Z+j$_{30}$ & Z+j$_{100}$ & $\mathrm{t\bar{t}}$
 \\ 
\midrule
LO & $16.93_{-3.22}^{+4.53}~\pb$ & $364.5_{-56.9}^{+53.2}~\pb$ & $63.77_{-6.52}^{+8.26}$~\pb & $5.258_{-0.479}^{+0.588}~\pb$ & $569.1_{-119.2}^{+163.5}$~\pb \\[6pt]
NLO & $39.13_{-6.62}^{+8.89}~\pb$ & $464.4_{-32.5}^{+19.7}~\pb$ & $88.16_{-5.68}^{+6.16}$~\pb & $9.254_{-0.798}^{+1.039}~\pb$ & $853.8_{-99.6}^{+99.7}$~\pb \\[6pt]
NNLO & $49.33_{-5.02}^{+4.41}~\pb$ & $470.9_{-6.8}^{+2.5}~\pb$ & $83.95_{-2.87}^{+0.49}$~\pb  &
$9.079_{-0.173}^{+0.026}$~\pb & $952.3_{-52.9}^{+33.1}$~\pb \\
\bottomrule
\end{tabular}  
\caption{Examples of uncertainty estimates from seven-point scale variations for hard processes in $pp$ collisions at $14~\TeV$ CM energy. Results for $t\bar{t}$ are obtained with \topxx~\cite{Czakon:2011xx}, all others with \nnlojet~\cite{NNLOJET:2025rno}.
The central scale choices are, respectively, $m_\mathrm{H}/2$, $m_\mathrm{Z}/2$, the geometric mean of $m_{T\mathrm{Z}}$ and $p_{T\mathrm{j}}$, $\,\sqrt{m_{TZ}\, p_{T\mathrm{j}}}=54$ and $120~\GeV$, and $m_t$.
}
\label{tab:sigmaNNLOjet}
\end{table}

\begin{figure}[tp]
\centering
\includegraphics*[width=0.9\textwidth]{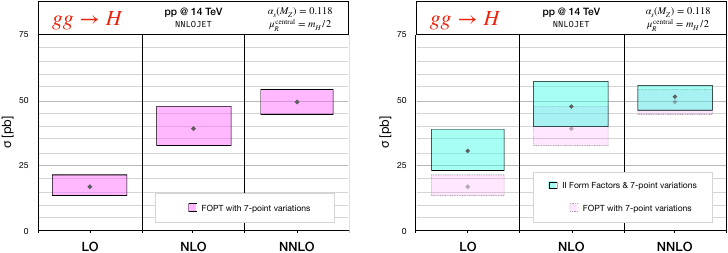}
\caption{Central cross sections and uncertainty bands for $gg\to H$ at LO, NLO, and NNLO, calculated with \nnlojet~\cite{NNLOJET:2025rno} for {\sl (left)} fixed-order perturbation theory with 7-point variations and {\sl (right)} the same with II form factors (and their uncertainty estimates) included. 
See \appRef{app:numbers} for a table of numerical values.\label{fig:plotH}}
\end{figure}
\begin{figure}[tp]
\centering
\includegraphics*[width=0.9\textwidth]{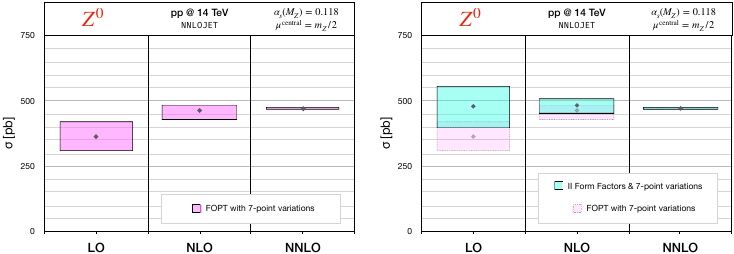}
\caption{Central cross sections and uncertainty bands for $Z^0$ production at LO, NLO, and NNLO, calculated with \nnlojet~\cite{NNLOJET:2025rno} for {\sl (left)} fixed-order perturbation theory with 7-point variations and {\sl (right)} the same with II form factors (and their uncertainty estimates) included. 
See \appRef{app:numbers} for a table of numerical values.\label{fig:plotZ}}
\end{figure}
\begin{figure}[tp]
\centering
\includegraphics*[width=0.9\textwidth]{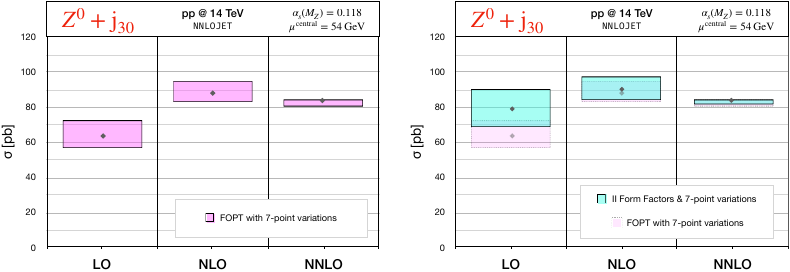}
\caption{Central cross sections and uncertainty bands for $Z+\mathrm{jet}$ at LO, NLO, and NNLO (with the jet defined by the anti-kT clustering algorithm~\cite{Cacciari:2008gp} with $\Delta R=0.4$ and $p_{T\mathrm{j}}\ge30\,\mathrm{GeV}$), calculated with \nnlojet~\cite{NNLOJET:2025rno} for {\sl (left)} fixed-order perturbation theory with 7-point variations and {\sl (right)} the same with II form factors (and their uncertainty estimates) included.
See \appRef{app:numbers} for a table of numerical values.\label{fig:plotZj30}}
\end{figure}
\begin{figure}[tp]
\centering
\includegraphics*[width=0.9\textwidth]{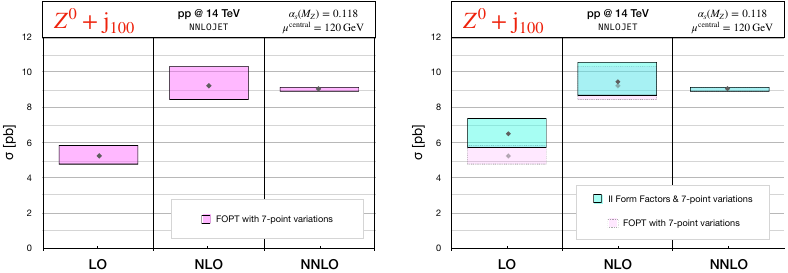}
\caption{Central cross sections and uncertainty bands for $Z+\mathrm{jet}$ at LO, NLO, and NNLO (with the jet defined by the anti-kT clustering algorithm~\cite{Cacciari:2008gp} with $\Delta R=0.4$ and $p_{T\mathrm{j}}\ge100\,\mathrm{GeV}$), calculated with \nnlojet~\cite{NNLOJET:2025rno} for {\sl (left)} fixed-order perturbation theory with 7-point variations and {\sl (right)} the same with II form factors (and their uncertainty estimates) included.
See \appRef{app:numbers} for a table of numerical values.\label{fig:plotZj100}}
\end{figure}

Any channels that do not involve colour annihilation are not
associated with any II form-factor uncertainty. Thus, e.g., processes
like DIS and VBF (at LC) do not acquire any uncertainty contribution
from this source.

\begin{figure}[tp]
\centering
\includegraphics*[width=0.9\textwidth]{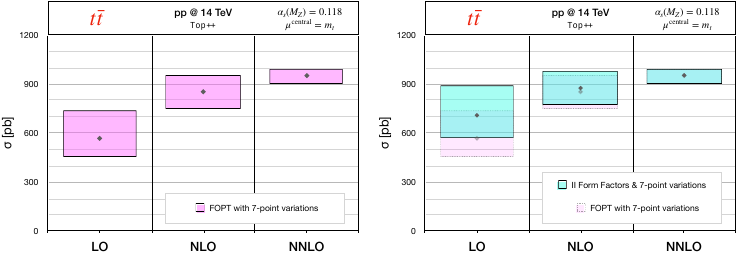}
\caption{Central cross sections and uncertainty bands for $t\bar{t}$ at LO, NLO, and NNLO, calculated with \topxx~\cite{Czakon:2011xx} for {\sl (left)} fixed-order perturbation theory with 7-point variations and {\sl (right)} the same with II form factors (and their uncertainty estimates) included.
See \appRef{app:numbers} for a table of numerical values.\label{fig:plotTT}}
\end{figure}

In \tabRef{tab:sigmaNNLOjet}, we compare cross sections for three of these processes, computed at increasing fixed order with \nnlojet~\cite{NNLOJET:2025rno} and \topxx~\cite{Czakon:2011xx}, with uncertainties from standard 7-point scale variations. The same numbers are illustrated in the left-hand panes of \figsRef{fig:plotH} -- \ref{fig:plotTT} (magenta bands). The right-hand panes of the same figures show the effects of including the full (exponentiated) II form factors (cyan bands), subtracting out the fixed-order contributions through the given order. This changes the central values upward. An additional uncertainty on the II form-factor contribution is also included, defined simply as half the size of the correction itself. This is then added in quadrature with the uncertainty coming from the 7-point scale variations (which is left unchanged).

For $gg\to H$, \figRef{fig:plotH}, the effect is dramatic especially for the first two orders. At LO, the effect is larger than the band covered by the LO scale variation, which is not surprising given the inadequacy of scale variations to provide a meaningful uncertainty for this process at LO. At NLO, the shift induced by the II form factor is comparable to the size of the scale variation, while the uncertainty band only becomes slightly larger\footnote{This is partly since we choose, arbitrarily, to add the 7-point  and II-form-factor uncertainties in quadrature.}. At NNLO, the II form-factor effect is smaller than the scale variation, and the uncertainty band remains almost the same, just with a slightly higher central value. 

For Drell-Yan, \figRef{fig:plotZ}, there is again a quite large effect at LO, with the central prediction moving outside the band covered by scale variations (which again are inadequate due to the absence of any powers of $\alpha_s$ in the Born-level process) to a value much closer to those calculated at higher orders, though with a large uncertainty. At NLO, the size of the upwards shift in the central value is again comparable to the size of the scale variation while the size of the estimated uncertainty band does not change dramatically. At NNLO, the Drell-Yan cross section is essentially unchanged. 

For $Z+\mathrm{jet}$ with $p_{T\mathrm{j}}\ge 30\,\mathrm{GeV}$, \figRef{fig:plotZj30}, pure fixed-order perturbation theory with 7-point variations (left pane) again exhibits the pattern of the LO and NLO bands not overlapping, while they do overlap once II form-factor corrections and their uncertainty estimates are included. The effect is most dramatic at LO, while already at NLO the total size of the uncertainty band only increases very slightly. 

For $Z+\mathrm{jet}$ with $p_{T\mathrm{j}}\ge 100\,\mathrm{GeV}$, \figRef{fig:plotZj100},
the absolute jump from LO to NLO in the pure fixed-order prediction (left-hand pane) is almost as dramatic as for Higgs production. (This appears consistent with the $K$ factors found for large $p_{\mathrm{T},j}$ in \cite{Gehrmann-DeRidder:2015wbt}.) Moreover, even though the Born-level process does now involve one power of $\alpha_s$, the 7-point variation is very small, and becomes larger at NLO. In the right-hand pane, we see that the estimated II form-factor effect, although it does change the LO cross-section estimate substantially, does not account for the full jump, nor does the estimated uncertainty overlap with the NLO bands. We attribute this to the LO calculation lacking a physical contribution which only appears at NLO and which neither scale variations nor II form factors account for: the contribution from hard Born-level dijet events radiating a comparatively soft $Z$ boson. We do not see this as invalidating a strategy of applying universal II form-factor corrections for colour-annihilation channels. It does, however, emphasise the important point that there can be other large effects in the perturbative series, which more work would be required to explore.

For $t\bar{t}$, \figRef{fig:plotTT}, the upwards shift in central value at LO is comparable to the size of the LO scale-variation band, while at higher orders the II form-factor effect is small.

\subsection{Top quark mass schemes}
\label{sec:masses}

In processes involving top quarks, the top-quark mass parameter, $m_t$, plays at least three important roles:
\begin{enumerate}
\item It sets the top-quark Yukawa coupling in top-Higgs interaction vertices. 
\item It appears in the propagators of virtual top-quark loops.
\item It sets the kinematically reconstructible invariant mass in top-quark decays. 
\end{enumerate}

Mathematically, $m_t$ is a renormalizable parameter of the Standard Model Lagrangian density, whose running is governed by renormalisation group evolution (RGE). When doing RGE-improved perturbation theory, questions therefore naturally arise as to which top-quark mass scheme (and scale) should be used to obtain a reasonable ``central prediction'' for each given application, and how to estimate robust uncertainties associated with this choice. The decision as to which renormalisation scheme to use is a choice, and should not have an impact on the level of physical observables. This statement is of course true up to the fact that we work in perturbation theory, and the conversion between different schemes is not known to all orders. Similarly to residual scale dependence in standard fixed-order calculations, the dependence of the results on the scheme choice is an artifact of truncation of higher-order terms in the perturbative expansion.

Here, we focus on questions associated with the first two roles above, in particular uncertainties associated with the choice between the $\overline{\mathrm{MS}}$ and the on-shell (OS) scheme, for processes involving virtual top quarks. 

\begin{figure}[t]
\begin{center}
\includegraphics[width=0.9\textwidth]{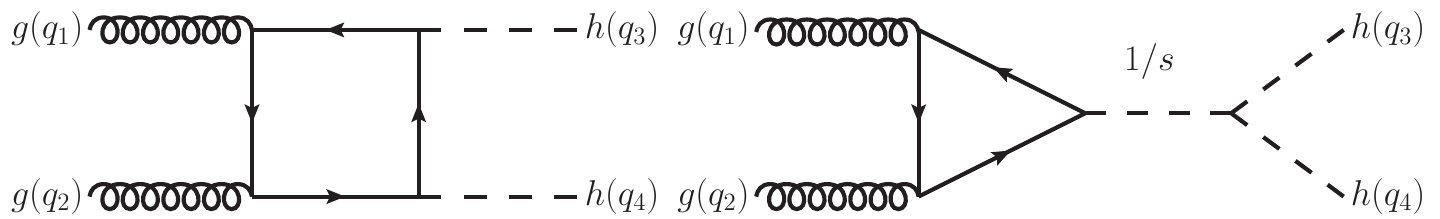}
\caption{\label{fig:hhdia} The leading order QCD diagrams contributing to the $gg \to HH$ process. The box diagram on the left-hand side is proportional to $y_t^2$, while the triangle diagram on the right-hand side is proportional to $y_t$.}
\end{center}
\end{figure}

A prime example of such a process is di-Higgs production in gluon fusion at the LHC, where at leading order the two external Higgs bosons are either both produced directly from attaching to the virtual top-quark loop, which we call {\emph{box}} type contributions, or a single off-shell Higgs boson is produced via an interaction with the top-quark loop and subsequently splits into two external Higgs bosons. The latter contributions we call {\emph{triangle}} type. See \figRef{fig:hhdia} for representative diagrams. Concretely, the $gg \rightarrow HH$ amplitude can be decomposed in terms of two form factors which have the following form
\begin{align}
&A_1 = T_F \frac{G_F}{\sqrt{2}} \frac{\alpha_s}{2 \pi} s\  
\left[ \frac{3 m_H^2}{s - m_H^2} A_{1,y_t \lambda_3} + A_{1, y_t^2} \right],&
&A_2 = T_F \frac{G_F}{\sqrt{2}} \frac{\alpha_s}{2 \pi} s\  
\left[ A_{2, y_t^2} \right].&
\label{eq:Ai}
\end{align}
Moreover, the form factors admit an expansion in powers of the strong coupling as
\begin{align}
A_{i,j} = \sum_{k=0} \left(\frac{\alpha_s}{2\pi} \right)^k A_{i,j}^{(k)}, \quad i=1,2,\ j=y_t^2, y_t \lambda_3,
\end{align}
where $j=y_t^2$ labels the box type diagrams, and $j=y_t \lambda_3$ the triangle type diagrams.
The results for the QCD corrections for this process containing the complete top-quark mass dependence are known only at next-to-leading order~\cite{Glover:1987nx,Borowka:2016ehy,Borowka:2016ypz,Baglio:2018lrj,Baglio:2020ini,Baglio:2020wgt,Bagnaschi:2023rbx,Campbell:2024tqg}. 
It was observed that the choice of the top-quark mass renormalisation scheme matters greatly for the observable distributions and since neither~OS nor~$\overline{\mathrm{MS}}$ scheme should be preferred a priori, it was suggested that the uncertainty should be determined by taking the envelope of the results obtained in the different schemes~\cite{Baglio:2020ini}. 

Currently, the choice of the top-quark mass renormalisation scheme is the largest source of theoretical uncertainty for this process~\cite{H:2026ksi}.
As higher-order calculations become available, the uncertainty due to this scheme choice will decrease, see for example partial results at next-to-next-to-leading order~\cite{Davies:2025ghl}, along with other sources of uncertainty. However, the progress to higher orders for processes with many scales is challenging. Therefore, it is interesting to investigate how much can be gained by considering the all-order structure of the top-quark mass corrections. 

The first steps in this direction were undertaken in \cite{Jaskiewicz:2024xkd}, where the focus was on understanding the high-energy behaviour of the $gg\to HH$ amplitude. Concretely, the region of interest has imposed the following scale hierarchy
\begin{equation}\label{eq:high-energy}
s, |t|, |u| \gg m_t^2 \gg m_H^2,
\end{equation}
where $s$, $t$, and $u$ are the usual Mandelstam invariants formed in scattering of two incoming particles into two outgoing ones. The scales $m_t$ and $ m_H$ are the top and Higgs particle masses, respectively. In the regime given by 
\eqRef{eq:high-energy}, the contributions due to the triangle type diagrams in \figRef{fig:hhdia} are power suppressed, and the dominant contribution originates from the box type diagrams. 

The investigation in~\cite{Jaskiewicz:2024xkd} begins with the observation that in the high-energy limit the result for one- and two-loop box contributions in the $\mathrm{OS}$ scheme can be written in the following way~\cite{Davies:2018qvx, Baglio:2018lrj, Baglio:2020ini},
\begin{align}
\label{eq:1.1}
A^{(0)}_{i,y_t^2} &= y_t^2\, f_i(s,t) 
+ \mathcal{O}(y_t^2 m_t^2)\,, \\ 
A^{(1)}_{i,y_t^2} &= 3 C_F \,A^{(0)}_i\,
\log\left[\frac{m_t^2}{s} \right]+ y_t^2 \, g_i(s,t) 
+ \mathcal{O}(y_t^2 m_t^2)\,,
\end{align}
where the functions $f_i$ are the leading order expressions for the form factors expanded to leading power in $m_t$. 
Their explicit form can be found in~\cite{Davies:2018qvx},
for example $f_1 = {8}/{s}$. 
The function $g_i(s,t)$ does not depend on $m_t$. 

The important observation is that the leading term proportional to $C_F$ at two loops can be traced back to the mass-renormalisation counter-term. 
Indeed, instead of the logarithm of $m_t^2/s$ in eq.~\eqref{eq:1.1}, converting the top-quark mass to the $\overline{{\rm{MS}}}$ scheme yields a logarithm of $\mu_t^2/s$~\cite{Baglio:2020ini}, with $\mu_t$ the $\overline{\mathrm{MS}}$ renormalisation scale.
Since the leading mass-dependent logarithm comes only from the mass-renormalisation counter-term, the full leading behaviour of $A^{(1)}_{i,y_t^2}$ in the high-energy limit can be predicted directly from the leading order $A^{(0)}_{i,y_t^2}$ result and the mass renormalisation counter-term.

The following questions which arise from this discussion that were answered in~\cite{Jaskiewicz:2024xkd} are: 
\begin{enumerate}
    \item Why is it that the box amplitudes have such a simple structure?
    \item Does this picture persist at higher orders in the strong coupling, or can super-leading or power-enhanced terms in the $m_t$ expansion start to enter in the integrals at higher loops  to spoil this structure?
\end{enumerate}

\begin{figure}
     \centering
     \begin{subfigure}[b]{0.48\textwidth}
         \centering
         \includegraphics[width=\textwidth]{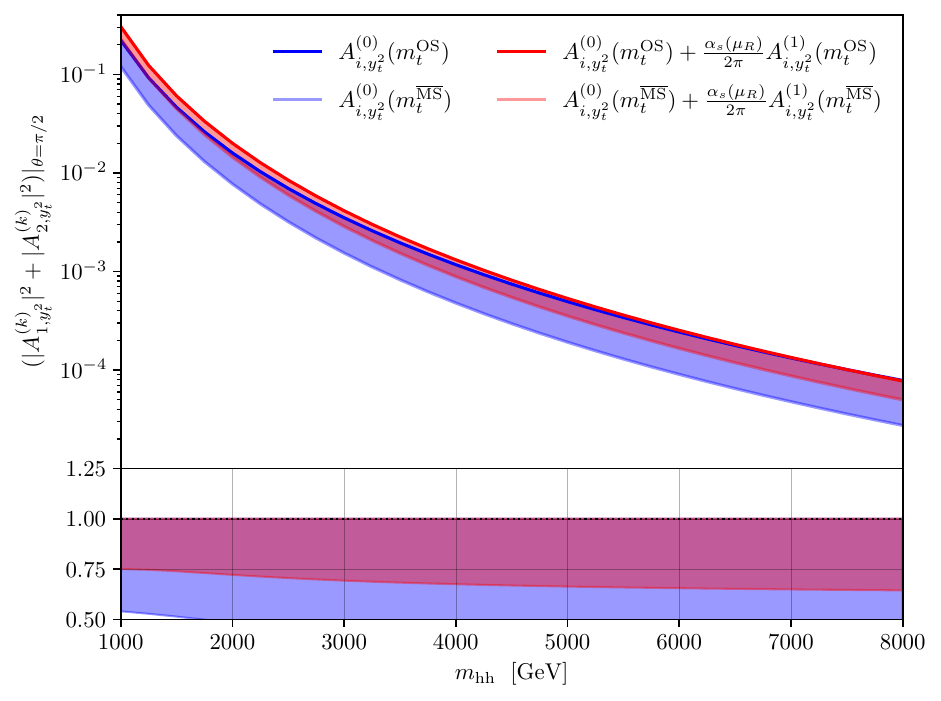}
         \caption{ }
         \label{sfig:all_sq_msbar_vs_os}
     \end{subfigure}
     \hfill
     \begin{subfigure}[b]{0.48\textwidth}
         \centering
         \includegraphics[width=\textwidth]{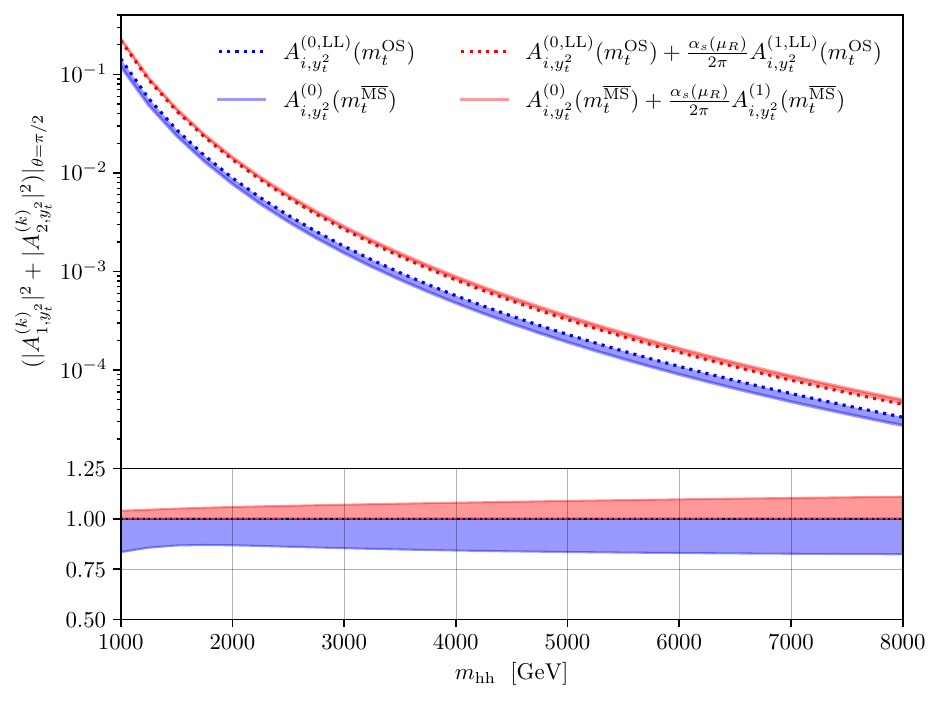}
         \caption{}
         \label{sfig:all_sq_msbar_vs_osll}
     \end{subfigure}
        \caption{Panel (a): Plot showing one- and two-loop squared form factors, where the top-quark mass is renormalised in either the ${\mathrm{OS}}$ and $\overline{\mathrm{MS}}$ scheme. Panel (b): The same quantities with the resummed tower of leading-power leading logarithms reinstated in the ${\mathrm{OS}}$ result.  A significant reduction in the size of the uncertainty is observed. Results shown are from Ref.~\cite{Jaskiewicz:2024xkd}.}
        \label{fig:all_sq_msbar_vs_osll}
\end{figure}

The analysis in~\cite{Jaskiewicz:2024xkd} consists of two main parts.
In the first, the Method of Regions (MoR)~\cite{Beneke:1997zp, Smirnov:1990rz, Smirnov:1994tg, Smirnov:2002pj} approach is used to perform a detailed fixed-order study to investigate the regions that contribute to the  scalar master integrals and to the full amplitude. 
In the second part, an effective field theory is constructed which formalises the findings from the method-of-regions analysis to all orders in $\alpha_s$.
The appropriate framework to capture the high-energy structure of the amplitude is Soft-Collinear Effective Field Theory (SCET)~\cite{Bauer:2000yr, Bauer:2001yt,Bauer:2002nz, Beneke:2002ph, Beneke:2002ni}, which is suitable for processes with multiple collinear directions.
What this analysis yields, is the identification of the origin of the large logarithms in $m_t^2/s$ in the $gg\to HH$ amplitudes at high-energies.  In the $\overline{\rm MS}$ scheme, the complete tower of these logarithms is contained implicitly in the top-quark mass evolution. However, it is truncated in the OS amplitude. In~\cite{Jaskiewicz:2024xkd} it was shown that this 
tower of logarithms can be reinstated in the OS result through 
\begin{align}
A^{(j,\,{\text{LL}})}_{i,y_t^2}(m_t^\mathrm{OS}) = \left(\frac{m^{\text{LL}}(\mu_t)}{m_t^\mathrm{OS}} \right)^2 A^{(j)}_{i,y_t^2}(m_t^\mathrm{OS}), \label{eq:a_ll}
\end{align}
where
\begin{align}\label{eq:schmeme-conv-all-order}
&m^{\text{LL}}(\mu) = {m_t^\mathrm{OS}} \exp \left[ a_{\gamma_m}^{\text{LL}}(\mu)\right]\, z_m({m_t^\mathrm{OS}}),&
&a_{\gamma_m}^{\text{LL}}(\mu) = \frac{3 C_F}{2 \beta_0} \ln \left(1 - \frac{\alpha_s(\mu)}{2 \pi} \beta_0 \ln \left(\frac{\mu^2}{{m_t^\mathrm{OS}}^2} \right) \right).&
\end{align}
The factor $z_m$ is the ratio of the OS and $\overline{\mathrm{MS}}$ renormalisation factors known to four loops~\cite{Chetyrkin:1999ys, Melnikov:2000qh,Marquard:2015qpa,Marquard:2016dcn} and $\beta_0 = 11 C_A/6 - 2/3 T_F n_f$. 
At the level of the $gg\to HH$ amplitude in the 
high-energy limit, the effect of including this tower of leading power leading logarithms in the OS
result is demonstrated in Fig.~\ref{fig:all_sq_msbar_vs_osll}. 
The key message is that the discrepancy between the two schemes obtained by a mismatch in the truncation order should not form part of the uncertainty budget, since in fact the leading logarithms (and beyond) in the conversion factor are {\emph{known}}. Therefore, they can be reinstated as above, which is shown to reduce the uncertainty between the two schemes~\cite{Jaskiewicz:2024xkd}. 
This analysis was recently extended to the cross-section level in~\cite{DelPio:2026cld}, following two approaches: implementing the LL resummation in the NLO $m_{HH}$ spectrum, and a dynamical top-mass scheme that uses the MSR mass near the $t\bar{t}$ threshold and the $\overline{\mathrm{MS}}$ mass at high $m_{HH}$. 
The latter yields a top-mass scheme uncertainty on the inclusive cross section at $\sqrt{s}=14$~TeV of ${}^{+5.2\%}_{-0.2\%}$, compared to ${}^{+5\%}_{-16\%}$ from the conventional OS--$\overline{\mathrm{MS}}$ envelope~\cite{H:2026ksi}.
One direction in improving the description is to work out the subleading power structure in $m_t^2/s$, which would extend the region of validity of the high-energy framework developed in~\cite{Jaskiewicz:2024xkd}. 
As a first step in this direction,~\cite{Jaskiewicz:2024xkd} develops the basis of subleading power operators. However, the factorisation structure at subleading powers is much richer than at the leading power, requiring for instance to treat endpoint divergences~\cite{Beneke:2019oqx,Beneke:2020ibj,Beneke:2022obx,Liu:2019oav,Bell:2022ott,Beneke:2026miy}. Therefore, it is interesting both from a theoretical and phenomenological standpoint to extend the framework of~\cite{Jaskiewicz:2024xkd} to next-to-leading power.
Moreover, it can also be useful to work on understanding the all-order structure of top-quark mass corrections in different expansion variables. For example, it could be insightful to consider the heavy-top limit, the threshold expansion at $s \sim 4 m_t^2$, or the small-$p_T$ expansion.

The SCET framework with massive quarks may also provide a useful starting point for understanding and resumming top-mass logarithms in H+Z, H+jet, and other loop-induced processes at high energies. A dedicated analysis is required in each case to establish whether the logarithms arise at leading power, as in HH production, or receive contributions at subleading power, potentially associated with soft-quark exchange. 
For H+Z and H+jet, the logarithms of $m_t$ have a very different origin than in the HH case. The same framework can be used to understand the origin of these and resum them; eventually (hopefully) reducing the uncertainty due to the top mass scheme. However, the details are very different and will require dedicated analysis.

In the context of the kinematically reconstructible top-quark mass, role 3 in the list at the beginning of this subsection, debate is ongoing on how to interpret the so-called Monte-Carlo (MC) mass (which is only defined implicitly through the numerical implementation of shower algorithms in MC event generators) and how this can be related to either the OS or $\overline{\mathrm{MS}}$ schemes. This remains an important open question but was not addressed during the discussions at the Aspen workshop.

\section{Non-gaussianity of theory uncertainties \label{sec:nonGauss}}

Even within the paradigm of conventional scale
variations, the numerical size of the variation itself is
ambiguous. As discussed in the introduction, the canonical choice is to vary the central scale choice(s) 
by a factor 2 in each direction. This is well motivated by logarithmic arguments but we cannot say with certainty which precise factor-$k$ variation would be equivalent
to a statistical $68\%$ (or 95\% etc.) confidence-level
interpretation. Moreover, there is no particular reason to assume the likelihood function for such variations to be  Gaussian.

These issues are of course not limited to scale variations but also
characterise many other uncertainty variations, spanning broad classes
of both theory uncertainties and experimental systematics, for which: 1) we
do not a priori know the precise size of the variation that most
closely could be said to correspond to a $68\%$ or $95\%$ etc.\ confidence
interval, and 2) there is often no fundamental reason for the
expectation to be Gaussian distributed. For instance, flatter
probability distributions could well be motivated, in 
cases where there is no strong preference for any particular value of
a given theory or systematic parameter.

Within a frequentist framework, situations of this kind have been considered
recently, in an approach referred to as ``errors on errors'' or, in more technical terminology, the Gamma Variance Model
(GVM)~\cite{Cowan:2018lhq,Canonero:2023sua,Canonero:2024kzk}. We
emphasise that it is not the objective of this framework to
provide a recipe for determining which variation is ``best'' (in the sense that it does not offer any guidance, say, on which scale variations should be done to estimate the theoretical uncertainty). It
simply allows to account for the fact that one has some uncertainty about what the actual uncertainty is, in a
statistically sound way and without needing to invoke Bayesian
methodologies~\cite{Cacciari:2011ze,Duhr:2021mfd}. (This does not 
imply any criticism of such methodologies.)

Naively, one might presume that introducing an uncertainty on an
uncertainty would simply inflate the width of a probability
distribution that would remain Gaussian in shape --- equivalent to the recipe
followed, e.g., by the PDG~\cite{ParticleDataGroup:2024cfk} when constructing averages for
measurements that are in tension with each other. But in the GVM, the
``error on the error'' changes not only the width but also the
nature of the probability distribution, to one with an almost Gaussian
core but wider tails. One practical consequence of this is that fits become
more stable against outliers, as demonstrated. e.g., in~\cite{Canonero:2023sua,Canonero:2024kzk}. Briefly explained, the GVM goodness-of-fit measure can explore further 
from the core of the distribution without being penalised quite as much as in the Gaussian case. Weighted averages then behave a bit more like medians than (Gaussian) means. This feature of the errors-on-errors
framework may also be of practical use in the context of allowing for
non-gaussianities in theory uncertainty estimates.

In the original works~\cite{Cowan:2018lhq,Canonero:2023sua,Canonero:2024kzk}, the context is mainly experimental measurements with the usual combination of statistical and systematic errors. The former can be treated within the conventional Gaussian modelling, while the latter typically represent best-effort estimates of various sources of systematics, whose true variances may not be precisely known. 
In the GVM, the best estimate of a systematic uncertainty is itself treated as uncertain, with its expectation given by the true systematic uncertainty and its relative uncertainty parametrised by a new statistical modelling parameter, $\epsilon$; for example, $\epsilon = 0.3$ corresponds to a 30\% relative uncertainty on the estimated systematic error.
Since this effectively allows for assigning uncertainties to the (systematic) uncertainties, this framework is also referred to as ``errors on errors''. 

This can be readily adapted to systematic uncertainties on theoretical calculations, such as those estimated by scale variations. Like in the case of experimental systematics, we know that scale variations themselves can at most be considered best-effort estimates. 
The GVM treats the nominal systematic uncertainty obtained from the scale variation as an uncertain estimate of the true systematic error. The parameter $\epsilon$ quantifies the relative uncertainty on this estimate. For $\epsilon = 0$, the quoted uncertainty is taken as exact and the usual Gaussian model is recovered, while $\epsilon > 0$ allows the true uncertainty to differ from its nominal estimate, resulting in increasingly non-Gaussian tails.

For definiteness, consider an observable (e.g., a cross section), $y$, whose central theoretical prediction is $y = y^\mathrm{central}_\mathrm{th}$. (Assume for simplicity here that the predicted value $y^\mathrm{central}_\mathrm{th}$ has negligble statistical uncertainty, since that can be dealt with via standard methods.) By performing a variation of a parameter in the underlying theoretical calculation, we have estimated a (single source of) systematic uncertainty and would normally quote the theoretical prediction as $y = y^\mathrm{central}_\mathrm{th} \pm \sigma_\mathrm{th}$. 

Within the Gaussian paradigm, that means that the log-likelihood of the true value of $y$ to be $y=\upsilon$ would contain a term
\begin{equation}
-\frac{(\upsilon - y^\mathrm{central}_\mathrm{th})^2}{2\sigma_\mathrm{th}^2}\,.
\end{equation}
In the GVM model, this is generalised to 
\begin{equation}
-\frac12\left(1 + \frac{1}{2\epsilon^2}\right)\ln\left(1 + 2\epsilon^2 \frac{(\upsilon - y^\mathrm{central}_\mathrm{th})^2}{v_\mathrm{th}}\right)\,,\label{eq:GVM}
\end{equation}
where $v_\mathrm{th}$ is the 
nominal estimate of the unknown true variance
\(\sigma_{\rm th}^2\).\footnote{The model is formulated
in terms of the variance estimate \(v_{\rm th}\), rather than the corresponding
standard deviation, for mathematical convenience. In particular, the variance
estimate is described by a Gamma distribution, leading to the analytic form
of the likelihood used here; see \cite{Cowan:2018lhq} for details.}
The parameter \(\epsilon\) quantifies the relative uncertainty on the estimated standard deviation
\(\sqrt{v_{\rm th}}\). 
Although this may look very different at first sight, one does recover the Gaussian behaviour for $\epsilon\to 0$, as can be verified, e.g., by doing a small-$\epsilon$ expansion of \eqRef{eq:GVM}:
\begin{eqnarray}
& \stackrel{\epsilon\to 0}{\to} & -\frac{(\upsilon - y^\mathrm{central}_\mathrm{th})^2}{2v_\mathrm{th}}\nonumber\\[2mm]
& & \,+\, \epsilon^2\left(\frac{(\upsilon - y^\mathrm{central}_\mathrm{th})^4}{2v_\mathrm{th}^2}
- \frac{(\upsilon - y^\mathrm{central}_\mathrm{th})^2}{v_\mathrm{th}}
\right) \nonumber\\[2mm]
& & \,+\,{\cal O}(\epsilon^4)~.
\end{eqnarray}
 For $\epsilon > 0$, the likelihood function develops wider-than-Gaussian tails. 
 In practice, \(\epsilon\)
need not be specified with high precision or fine granularity. It is more
naturally regarded as an approximate scale, with values such as
\(\epsilon=0,\,0.25,\,0.5,\,0.75,\) or \(1\) chosen according to the perceived
reliability of the estimated uncertainty. Such assignments are necessarily
approximate, but already provide additional flexibility compared with the
fixed-variance assumption \(\epsilon=0\). Accounting for uncertainty in the
estimated error by choosing an $\epsilon>0$ extends the confidence intervals to larger values of \((\upsilon-y_{\rm th}^{\rm central})^2\), with higher
confidence intervals (e.g. 95\% or 99\%) inflated by larger factors than lower
ones.

%

As a rule of thumb,
values of $\epsilon \lesssim0.1$ do not produce substantial
changes relative to the ordinary Gaussian framework, while substantial deviations typically become visible for $\epsilon \gtrsim 0.3$. Values of epsilon around 1 correspond to an uncertainty on the error estimate comparable to the estimate itself, while $\epsilon > 1$ represents more than 100\% relative uncertainty on the estimated error.

In the case of
scale variations, it may not be unreasonable to allow for fairly
large $\epsilon$ values, since
we are typically uncertain about whether the true error could differ from the estimated one by more than 10\%. 

Ultimately, it would be useful to collect some  
experience by exploring and comparing different options in practical cases. For interpretations, it may also need a little getting-used-to, to gain intuition for the implications of the non-Gaussian relationships between the
various confidence-level contours.  For improvements to the accuracy of the statistical inference, see~\cite{Canonero:2023sua}; for generalisations to correlated uncertainties, see~\cite{Canonero:2024kzk}.

\section{Summary}
\label{sec:summary}

Conventional 7-point renormalisation- and factorisation-scale variations frequently fail to capture the true size of missing higher-order corrections. This is especially acute for the LO/NLO(+PS) accuracy at which most present-day Monte-Carlo event generators --- and hence much of current LHC Run-3 phenomenology --- operate.

Motivated by this, we examined three concrete and theoretically well-understood sources of higher-order terms that 7-point scale variations are structurally blind to, with the aim of turning them into usable, quantitative uncertainty estimates within a standard frequentist framework: 1) scale variations for multi-scale problems, 2) colour-annihilation form factors, and 3) top-quark mass scheme variations. 

Ultimately, the ideal approach to tackling missing-higher-order uncertainties is of course to compute more. For full-fledged experimental and phenomenological applications, this also necessitates the development of parton-shower algorithms that can reach higher formal accuracies, and which can match fixed-order singularity structures at high orders, along with formal matching schemes that enable rigorous combinations of such algorithms with fixed-order calculations beyond (N)NLO. This remains a high-priority long-term program. 

We also summarised the Gamma Variance Model (``errors on errors''), a frequentist framework that allows for deviations from Gaussian uncertainties by modelling the uncertainties themselves as random variates, with a degree of uncertainty quantified by a new statistical modelling parameter $\epsilon$.

We hope this summary will be useful in the context of the ongoing discussion on how to improve the coverage and rigour of perturbative uncertainty estimates.

\subsection*{Acknowledgements}
This work was initiated at the Aspen Center for Physics, which is supported by National Science Foundation Grant PHY-2210452. We thank F.~Febres-Cordero and B.~Mistlberger for co-organising the workshop in Aspen, and for helpful comments on this work. We are also grateful to M.~Neubert for insightful discussions in Aspen on II form factors. MK is supported via the PAPIIT (IA103126) and  SECIHTI (CBF-2025-I-615) grants. DR is supported by the STFC under grant agreement ST/P006744/1 and by the European Union under the HORIZON program in Marie Sk{\l}odowska-Curie project No. 101153541. PS is supported by ARC grant DP220103512. RS is supported by the United States Department of Energy under Grant Contract DE-SC0012704.

\appendix
\section{Cross sections with II form factors}
\label{app:numbers}
\begin{table}[h!]
\centering
\begin{tabular}{lccc}
\toprule
\color{red}$gg\to H$& ~~$\sigma_{\mathrm{FO}}$ [pb]~~ & ~~$\Delta\sigma_\mathrm{II}$ [pb]~~ & ~$\sigma_{\mathrm{FO}} + \Delta\sigma_{\mathrm{II}} $ [pb] \\
\midrule
LO 
& $16.93_{-3.22}^{+4.53}$ 
& $13.70_{-6.85}^{+6.85}$ 
& $30.63_{-7.57}^{+8.21}$\\[6pt]
NLO 
& $39.13_{-6.62}^{+8.89}$ 
& $8.46_{-4.23}^{+4.23}$ 
& $47.59_{-7.86}^{+9.85}$\\[6pt]
NNLO 
& $49.33_{-5.02}^{+4.41}$ 
& $2.00_{-1.00}^{+1.00}$ 
& $51.33_{-5.12}^{+4.52}$\\
\bottomrule
\end{tabular}
\caption{Fixed-order cross sections and 7-point variations (first column), the estimated all-orders II form-factor correction (second column), and combined estimated cross section with the scale-variation and II-form-factor uncertainties summed in quadrature (last column), for $gg\to H$ at the LHC at $\sqrt{s}=14\,\mathrm{TeV}$, with $\mu^\mathrm{central}=m_H/2=60\,\mathrm{GeV}$.
These numbers correspond to the plots in \figRef{fig:plotH}.}
\end{table}

\begin{table}[h!]
\centering
\begin{tabular}{lccc}
\toprule
\color{red}$Z$& ~~$\sigma_{\mathrm{FO}}$ [pb]~~ & ~~$\Delta\sigma_\mathrm{II}$ [pb]~~ & ~$\sigma_{\mathrm{FO}} + \Delta\sigma_{\mathrm{II}} $ [pb] \\
\midrule
LO 
& $364.5_{-56.9}^{+53.2}$
& ~~$115.6^{+57.8}_{-57.8}$~~
& $480.1^{+78.56}_{-81.1}$\\[6pt]
NLO 
& $464.4_{-32.5}^{+19.7}$ 
& $19.4_{-9.7}^{+9.7}$ 
& $483.8_{-33.9}^{+21.9}$\\[6pt]
NNLO 
& $470.9_{-6.8}^{+2.5}$ 
& $1.8_{-0.9}^{+0.9}$ 
& $472.7_{-6.9}^{+2.7}$\\
\bottomrule
\end{tabular}
\caption{Fixed-order cross sections and 7-point variations (first column), the estimated all-orders II form-factor correction (second column), and combined estimated cross section with the scale-variation and II-form-factor uncertainties summed in quadrature (last column), for $Z$ production at the LHC at $\sqrt{s}=14\,\mathrm{TeV}$, with $\mu^\mathrm{central}=m_Z/2=45.6\,\mathrm{GeV}$. 
These numbers correspond to the plots in \figRef{fig:plotZ}.}
\end{table}

\begin{table}[ht!]
\centering
\begin{tabular}{lccc}
\toprule
\color{red}$Z+\mathrm{j}_{30}$& ~~$\sigma_{\mathrm{FO}}$ [pb]~~ & ~~$\Delta\sigma_\mathrm{II}$ [pb]~~ & ~$\sigma_{\mathrm{FO}} + \Delta\sigma_{\mathrm{II}} $ [pb] \\
\midrule
LO 
& $63.77_{-6.52}^{+8.26}$
& ~~$15.42^{+7.71}_{-7.71}$~~
& $79.19^{+11.29}_{-10.10}$\\[6pt]
NLO 
& $88.16_{-5.68}^{+6.16}$ 
& $2.22_{-1.11}^{+1.11}$ 
& $90.38_{-5.79}^{+6.26}$\\[6pt]
NNLO 
& $83.95_{-2.87}^{+0.49}$
& $0.15_{-0.08}^{+0.08}$
& $84.10_{-2.87}^{+0.50}$\\
\bottomrule
\end{tabular}
\caption{Fixed-order cross sections and 7-point variations (first column), the estimated all-orders II form-factor correction (second column), and combined estimated cross section with the scale-variation and II-form-factor uncertainties summed in quadrature (last column), for $Z+\mathrm{jet}$ production at the LHC at $\sqrt{s}=14\,\mathrm{TeV}$, with $\mu^\mathrm{central}=120\,\mathrm{GeV}$ and $p_{T\mathrm{j}}\ge 30\,\mathrm{GeV}$ (anti-kT, $\Delta R=0.4$). 
These numbers correspond to the plots in \figRef{fig:plotZj30}.}
\end{table}

\begin{table}[ht!]
\centering
\begin{tabular}{lccc}
\toprule
\color{red}$Z+\mathrm{j}_{100}$& ~~$\sigma_{\mathrm{FO}}$ [pb]~~ & ~~$\Delta\sigma_\mathrm{II}$ [pb]~~ & ~$\sigma_{\mathrm{FO}} + \Delta\sigma_{\mathrm{II}} $ [pb] \\
\midrule
LO 
& $5.26_{-0.48}^{+0.59}$
& ~~$1.26^{+0.63}_{-0.63}$~~
& $6.52^{+0.86}_{-0.79}$\\[6pt]
NLO 
& $9.25_{-0.80}^{+1.04}$ 
& $0.23_{-0.12}^{+0.12}$ 
& $9.48_{-0.81}^{+1.05}$\\[6pt]
NNLO 
& $9.08_{-0.17}^{+0.03}$
& $0.02_{-0.01}^{+0.01}$
& $9.09_{-0.17}^{+0.03}$\\
\bottomrule
\end{tabular}
\caption{Fixed-order cross sections and 7-point variations (first column), the estimated all-orders II form-factor correction (second column), and combined estimated cross section with the scale-variation and II-form-factor uncertainties summed in quadrature (last column), for $Z+\mathrm{jet}$ production at the LHC at $\sqrt{s}=14\,\mathrm{TeV}$, with $\mu^\mathrm{central}=120\,\mathrm{GeV}$ and $p_{T\mathrm{j}}\ge 100\,\mathrm{GeV}$ (anti-kT, $\Delta R=0.4$). 
These numbers correspond to the plots in \figRef{fig:plotZj100}.}
\end{table}

\begin{table}[ht!]
\centering
\begin{tabular}{lccc}
\toprule
\color{red}$t\bar{t}$& ~~$\sigma_{\mathrm{FO}}$ [pb]~~ & ~~$\Delta\sigma_\mathrm{II}$ [pb]~~ & ~$\sigma_{\mathrm{FO}} + \Delta\sigma_{\mathrm{II}} $ [pb] \\
\midrule
LO 
& $569.1_{-119.2}^{+163.5}$
& $140.8^{+70.4}_{-70.4}$
& $709.9^{+178.0}_{-138.4}$\\[6pt]
NLO 
& $853.8_{-99.6}^{+99.7}$ 
& $22.5_{-11.2}^{+11.2}$ 
& $876.3_{-100.2}^{+100.3}$\\[6pt]
NNLO 
& $952.3_{-52.9}^{+33.1}$ 
& $1.8_{-0.9}^{+0.9}$ 
& $954.1_{-52.9}^{+33.1}$\\
\bottomrule
\end{tabular}
\caption{Fixed-order cross sections and 7-point variations (first column), the estimated all-orders II form-factor correction (second column), and combined estimated cross section with the scale-variation and II-form-factor uncertainties summed in quadrature (last column), for $t\bar{t}$ production at the LHC at $\sqrt{s}=14\,\mathrm{TeV}$, with $\mu^\mathrm{central}=m_t$. 
These numbers correspond to the plots in \figRef{fig:plotTT}.}
\end{table}

\clearpage
\bibliographystyle{SciPost_bibstyle}
\bibliography{main}
\end{document}